\documentstyle{elsart}

\newcommand{\Mpc}{$h^{-1}$~{\rm Mpc}}
\newcommand{\hmpc}{$h$~{\rm Mpc$^{-1}$}}
\newcommand{\apj}{{\em Astrophys. J.}} 
\newcommand{\aj}{{\em Astron. J.}} 
\newcommand{\mn}{{\em Mon. Not. R. astr. Soc.}}
\newcommand{\aanda}{{\em Astron. Astrophys.}}
\newcommand{\nat}{{\em Nature}}
\newcommand{\apjs}{{\em Astrophys. J. Suppl.}} 

\begin{document}
\begin{frontmatter}
\title{ Large Scale Structure}

\author{Jaan Einasto}

\address{Tartu Observatory, EE-61602 T\~oravere, Estonia}


\begin{abstract}

I give a review of catalogues of galaxies, clusters of galaxies and
superclusters -- sources of information to study the large--scale
structure of the Universe.  Thereafter I shall discuss the power
spectrum of density perturbations, and the correlation function --
principal description functions which characterize the large--scale
structure. I shall pay special attention to the geometric
interpretation of these functions, i.e.\ to the way in which the
various properties of the distribution of galaxies in systems and
systems themselves are reflected in these functions.  Finally, a
discuss cosmological parameters which characterize general properties
of the Universe -- the Hubble constant, densities of various
populations of the Universe, and parameters of the power spectrum of
galaxies and matter.

\end{abstract}
\end{frontmatter}

  
\section{ Introduction}

In this review I shall accept certain paradigms on the structure of
the Universe.  Paradigms on the structure of the Universe have evolved
considerably during the last hundred years.  This process is
continuing until the present time, and changes occur quite often, so
it is important that we clearly state the present paradigms.  I shall
use the term ``Universe'' for the real world around us, and the term
``universe'' for a model of the Universe (say Friedmann--universe).
Our accepted paradigms are:

\begin{itemize}
\item{} The Universe evolves from an explosive event termed ``Big
Bang'', through the inflation, the radiation--domination era to the
matter--domination era; probably we live now in the next era where the
dominating constituent of the Universe is dark energy.

\item{} The principal force driving the cosmological evolution is gravity. 

\item{} Density perturbations grow from small random fluctuations
generated during the inflation; perturbations have Gaussian
distribution.

\item{} The main constituents of the Universe are: baryonic matter
(stars and planets, hot and cold gas, and primordial gas in voids),
dark matter, either cold (CDM) or hot (HDM), and dark (vacuum) energy.
 
\item{} The Universe is flat -- the total mean density of all its
populations is equal to the critical density.

\end{itemize}
 
There exist a number of excellent reviews on the subject
``Large--scale structure of the Universe'', the most recent one with
references to earlier work is the talk by Guzzo \cite{g99} in the
19$^{th}$ Texas Symposium.  The term ``Large--scale structure of the
Universe'' itself originated in the contemporary meaning as the title
of an IAU Symposium \cite{le78}, where the presence of filamentary
distribution of galaxies and clusters with large empty voids between
them was first reported.  In this review I use the experience
collected at Tartu Observatory during the last 25 -- 30 years of the
study of the Universe.  I shall pay attention to cosmographic aspects
of the problem, in particular to the geometrical and physical
interpretation of descriptive functions.  Also I try to show how
advances in observational cosmology have changed our theoretical
understanding of the formation and evolution of the structure of the
Universe.

\section{Catalogues of galaxies, clusters and superclusters}

Our understanding of the structure of the Universe is based on the
distribution of galaxies.  Until the mid--1970s the number of galaxies
with known distances (redshifts) was very small, thus conclusions on
the structure were based on counts of galaxies.  The largest of such
counts was compiled in Lick Observatory by Shane \& Virtanen
\cite{sv67}.  This catalogue was analyzed by Seldner \etal\
\cite{ssgp77} and played a crucial role in the development of the
hierarchical clustering scenario of structure formation by Peebles
\cite{p80}.

A big step in the study of the clustering of galaxies and clusters of
galaxies was made by visual inspection of the Palomar Observatory Sky
Survey plates with the aim to produce catalogues of galaxies and
clusters of galaxies.  The first of these catalogues was prepared by
Abell \cite{a58} for clusters of galaxies.  This catalogue covers the
sky north of declination $-27^{\circ}$.  Abell, Corwin \& Olowin
\cite{aco} extended the cluster catalogue to the southern sky. Both
these catalogues together contain 4074 clusters.  A much larger
catalogue was compiled by Zwicky \etal\ \cite{z}; in this catalogue
all galaxies brighter than photographic magnitude $m_{ph}\simeq 15.7$
as well as clusters of galaxies north of declination $-2.5^{\circ}$
are listed.  Abell and Zwicky used rather different definitions of
clusters.  Abell clusters contain at least 30 galaxies in a magnitude
interval of $\Delta m =2$, starting from the third brightest galaxy,
and located within a radius of 1.5~\Mpc\ (we use in this paper the
Hubble constant in units $H_0 = 100~h$ km~s$^{-1}$~Mpc$^{-1}$).
Distances of clusters were estimated on the basis of the brightness of
the 10th brightest galaxy.  Clusters were divided to richness and
distance classes.  Zwicky used a more relaxed cluster definition, with
at least 50 galaxies in a magnitude interval of $\Delta m =3$,
starting from the brightest galaxy, located within a contour where the
surface density of galaxies exceeded a certain threshold.  Due to
these differences some Zwicky clusters are actually central parts of
superclusters which contain several Abell clusters and groups of
galaxies (an example is the Perseus cluster).  Since the definition of
clusters in the Abell catalogue is more exact, this catalogue has
served for a large number of studies of the structure of the Universe.
On the other hand, the Zwicky catalogue of galaxies was the basic
source of targets for redshift determinations.

An early catalogue of bright galaxies was compiled by Shapley \& Ames
\cite{sa32}.  Sandage \& Tammann \cite{st81} published a revised
version of this catalogue; it contains data on galaxies brighter than
13.5 magnitude, including redshifts.  This catalogue, and the
compilation of all available data on bright galaxies by de
Vaucouleurs, de Vaucouleurs, \& Corwin \cite{ref76} were the sources
of distances which allowed to obtain the first 3--dimensional distributions
of galaxies.  Much more detailed information on the spatial
distribution of galaxies was obtained on the basis of redshifts,
measured at the Harvard Center for Astrophysics (CfA) for all Zwicky
galaxies brighter than $m_{ph}=14.5$.  Later this survey was extended
to galaxies brighter than $m_{ph}=15.5$ (the second CfA catalogue),
and to galaxies of the southern sky (Southern Sky Redshift
Survey)\cite{DaCosta99}.

These early redshift compilations made it possible to discover the
filamentary distribution of galaxies and clusters forming huge
superclusters, as well as the absence of galaxies between them.  These
results were first reported in the IAU Symposium on Large--Scale
Structure of the Universe \cite{je78,tetal78,tg78,tf78} and
demonstrated that the pancake scenario of structure formation by
Zeldovich \cite{z70,z78} fits observations better than the
hierarchical clustering scenario.  More detailed studies of the
structure formation by numerical simulations showed that the original
pancake scenario by Zeldovich also has weak points -- there is no fine
structure in large voids between superclusters observed in the real
Universe \cite{zes82} and the structure forms too late \cite{dp83},
thus a new scenario of structure formation was suggested based on the
dominating role of the cold dark matter in structure evolution
\cite{bfpr84}.  In a sense the new scenario is a hybrid between the
original Peebles and Zeldovich scenarios: structure forms by
hierarchical clustering of small structures within large filamentary
structures -- superclusters.

The next big step in the study of the large--scale distribution of
galaxies was made on the basis of the catalogue of galaxies formed on
the basis of digitized images of the ESO Sky Survey plates using the
Automated Plate Measuring (APM) Facility \cite{m90,m97}.  The APM
galaxy catalogue covers 185 ESO fields, is complete up to magnitude
$b_j = 20.5$, and was the basis for a catalogue of clusters prepared
by Dalton \etal\ \cite{dm97}.  The analysis of the APM galaxy sample
showed that properties of the distribution of galaxies differ from the
standard CDM model which assumed that the density of matter is equal
to the critical density. A low--density model with cosmological term
(dark energy) fits the data better \cite{eetal}.

The modern era of galaxy redshift catalogues started with the Las
Campanas Redshift Survey (LCRS).  Here, for the first time,
multi--object spectrographs were used to measure simultaneously
redshifts of 50 -- 120 galaxies \cite{LCRS96}.  The LCRS covers 6
slices of size $1.5 \times 80$ degrees, the total number of galaxies
with redshifts is $\sim 26,000$, and the limiting magnitude is $b_j =
18.8$.  Presently several very large programs are under way to
investigate the distribution of galaxies in a much larger volume.  The
largest project is the Sloan Digital Sky Survey (SDSS), a cooperative
effort of several North--American institutions with participants from
Japan \cite{sdss}.  This survey covers the whole northern sky and a
strip in the southern sky. The sky is first imaged in five photometric
bands to a limiting magnitude about 23 (the limit varies with spectral
bands), thereafter redshifts are measured for all galaxies up to a
magnitude $\sim 18$, and active galactic nuclei (AGN) up to $\sim 19$;
additionally a volume--limited sample of redshifts of bright
elliptical galaxies is formed. The total number of galaxies with
measured redshifts will probably exceed one million.  Another large
redshift survey uses the 2--degree--Field \cite{2dF} spectrograph of
the Anglo--Australian Telescope.  This survey is based on the APM
galaxy catalogue and covers two large areas of size $75^{\circ} \times
12.5^{\circ}$ and $65^{\circ} \times 7.5^{\circ}$ with limiting
magnitude $b_j \approx 19.5$.  The goal is to measure about 250,000
redshifts.  It is expected that new redshift surveys give us the
possibility to investigate the detailed structure of the Universe up
to a distance of $\approx 2000$~\Mpc.

The largest systems of galaxies are superclusters, which are defined
as the largest systems of galaxies and clusters still isolated from
each other.  Catalogues of superclusters have been constructed using
Abell clusters of galaxies. The latest compilation by Einasto \etal\
\cite{e97c} contains 220 superclusters with at least two member
clusters.

\section{Distribution of galaxies and clusters}

\begin{figure}[ht]
\vspace*{16.5cm} 
\caption{The distribution of galaxies (upper panel), and particles 
in a 2-D simulation (lower panel). For explanations see text.}
\includegraphics{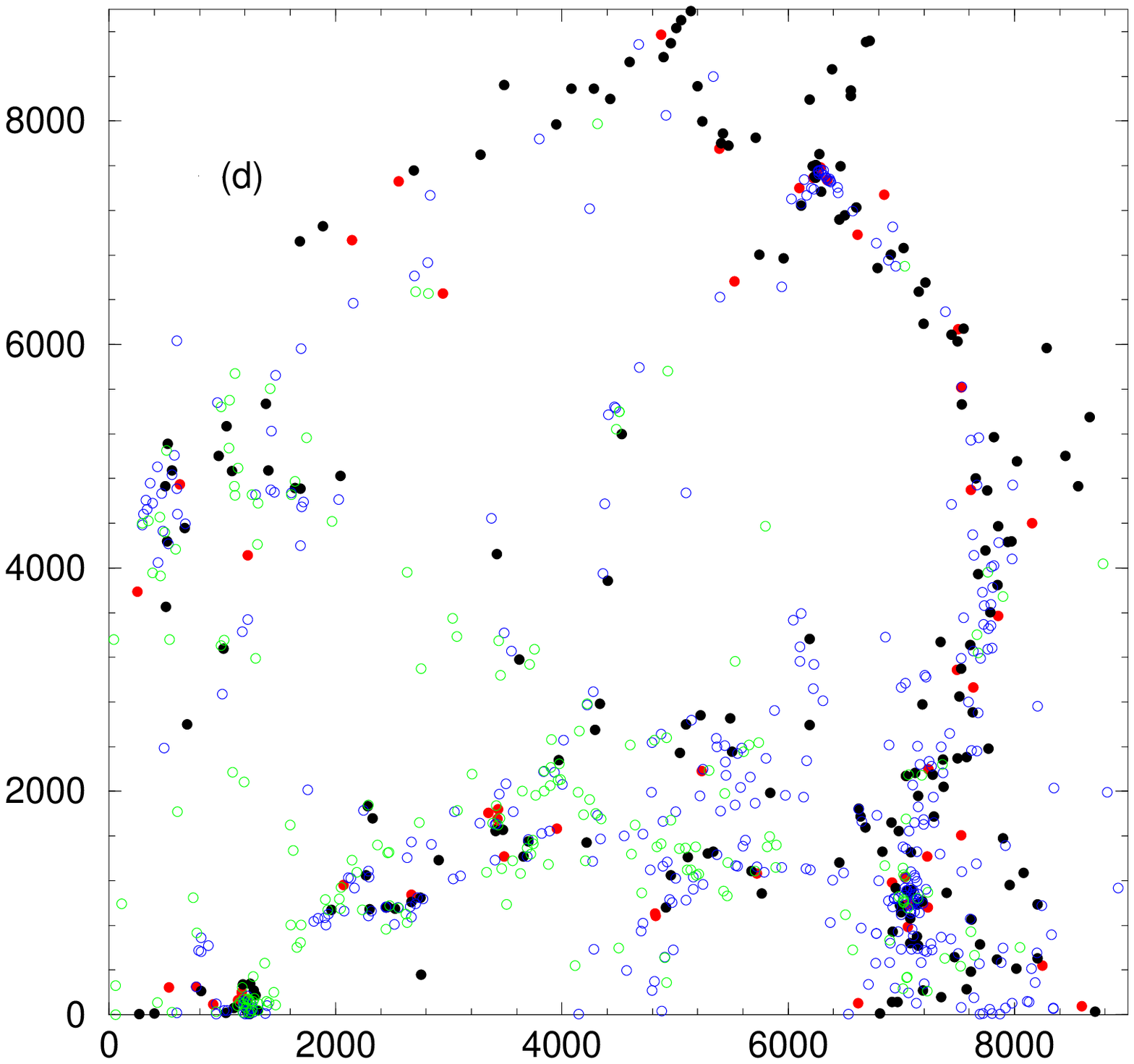} 
\includegraphics{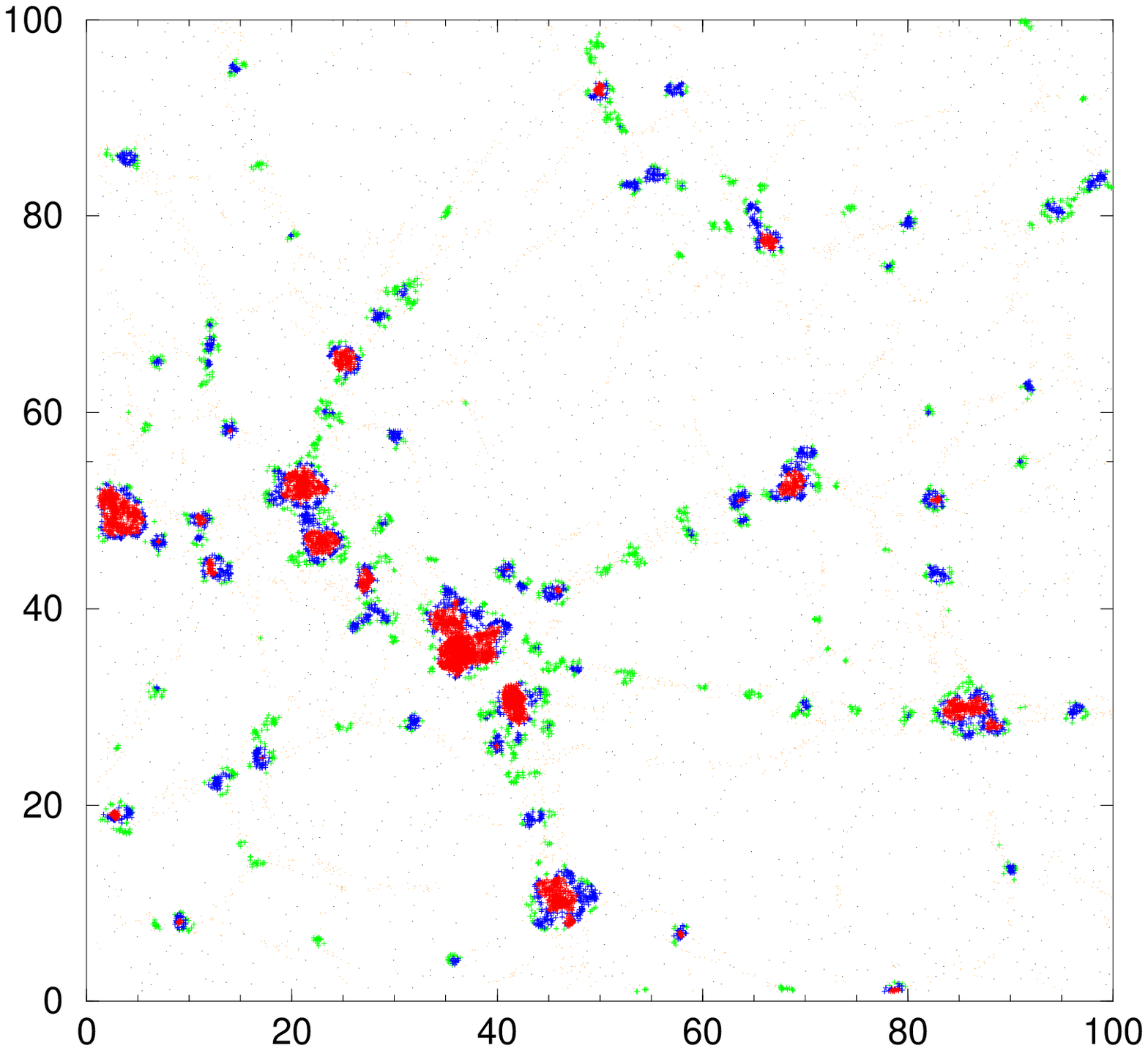}
\label{fig:gal}
\end{figure}

In Figure~\ref{fig:gal} I show the distribution of galaxies of various
luminosity in a volume--limited sample through the Virgo, Coma and
Hercules superclusters.  We use supergalactic coordinates $Y$ and $Z$
in km/s, respectively, in a sheet $0 \leq X < 10$~\Mpc.  Bright
galaxies ($M_B \leq -20.3$) are plotted as red dots, galaxies $-20.3 <
M_B \leq -19.7$ as black dots, galaxies $-19.7 < M_B \leq -18.8$ as
open blue circles, galaxies $-18.8 < M_B \leq -18.0$ as green circles
(absolute magnitudes correspond to Hubble parameter $h=1$).
High-density regions are the Local, the Coma and the Hercules
superclusters in the lower left, lower right and upper right corners,
respectively. The long chain of galaxies between Coma and Hercules
superclusters is called the Great Wall. Actually it is a filament.
For comparison, the distribution of particles in a 2--dimensional
simulation is also plotted in a box of side-length 100~\Mpc.
Different colors indicate the density value of the particle
environment.  Particles in voids (density $\varrho <1$) are shown as
black dots; particles in the density interval $1 \leq \varrho < 5$
form filaments of galaxies (orange dots); particles with densities $5
\leq \varrho < 10$ (green dots) form groups of galaxies; particles
with $10 \leq \varrho < 20$ (blue dots) form clusters; and particles
with $\varrho \geq 20$ (red dots) are in very rich clusters.
Densities are expressed in units of the mean density in the
simulation; they are calculated using a smoothing length of
1~\Mpc. Three--dimensional simulations have similar behaviour. This
Figure emphasizes that particles in high-density regions simulate
matter associated with galaxies, and that the density of the particle
environment defines the type of the structure. In both Figures we see
the concentration of galaxies or particles to clusters and filaments,
and the presence of large under-dense regions.  There exists, however,
one striking difference between the distribution of galaxies and
simulated particles -- there is a population of smoothly distributed
particles in low-density regions in simulations, whereas in the real
Universe voids are completely empty of any visible matter.  This
difference is due to differences of the evolution of matter in under--
and over--dense regions.

\begin{figure*}[ht]
\vspace*{5.5cm}
\caption{Left: Density evolution in over-- and under--dense regions
(solid and dashed lines, respectively) for two epochs of caustics
formation.  Right: Density perturbations of various
wavelengths. Under--dense regions ($D<1$) become voids; strongly
over--dense regions ($D>1.3$) -- superclusters (cluster chains);
moderately over--dense regions ($1 < D < 1.3$) -- filaments of groups
and galaxies.}
\includegraphics{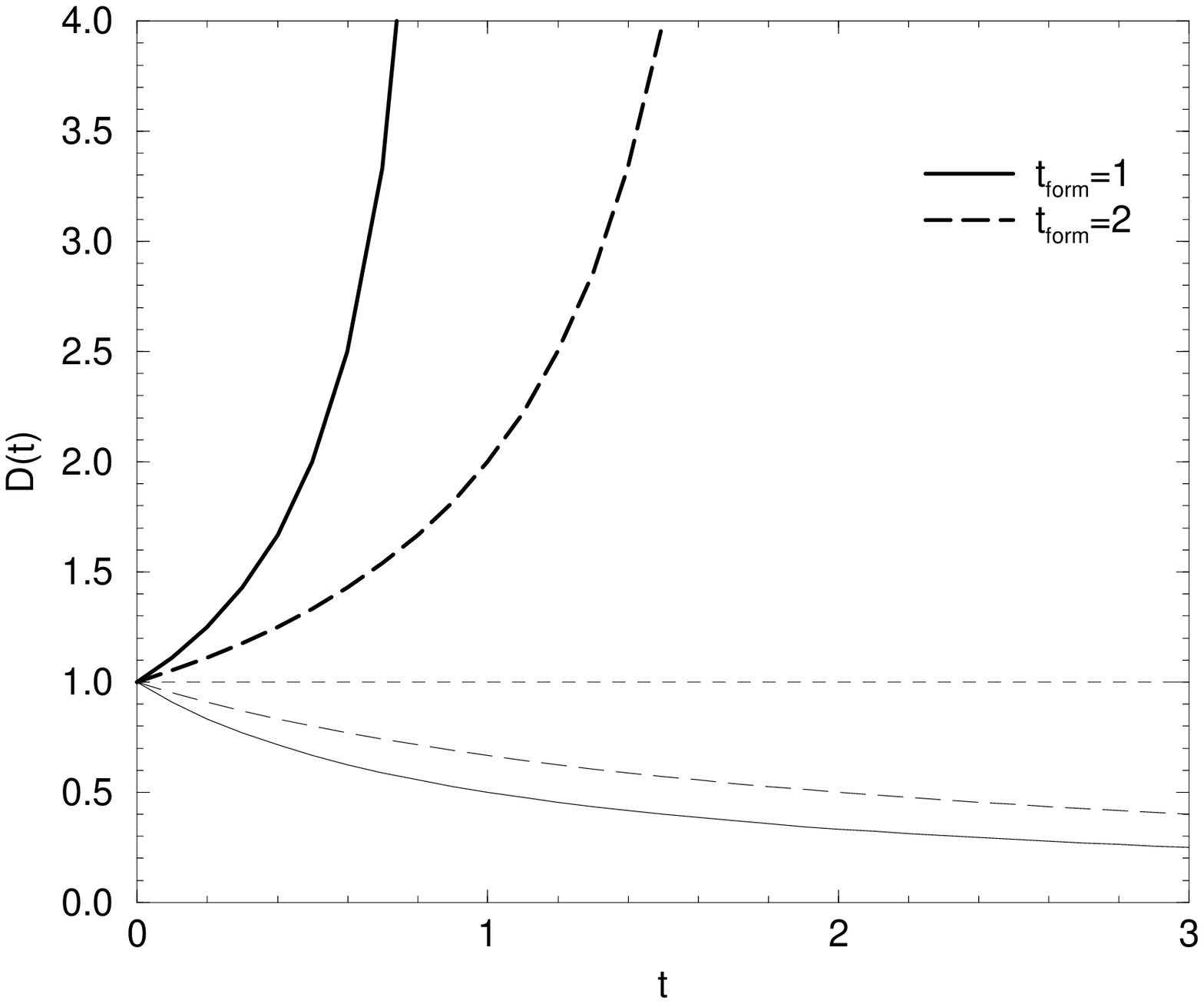}
\includegraphics{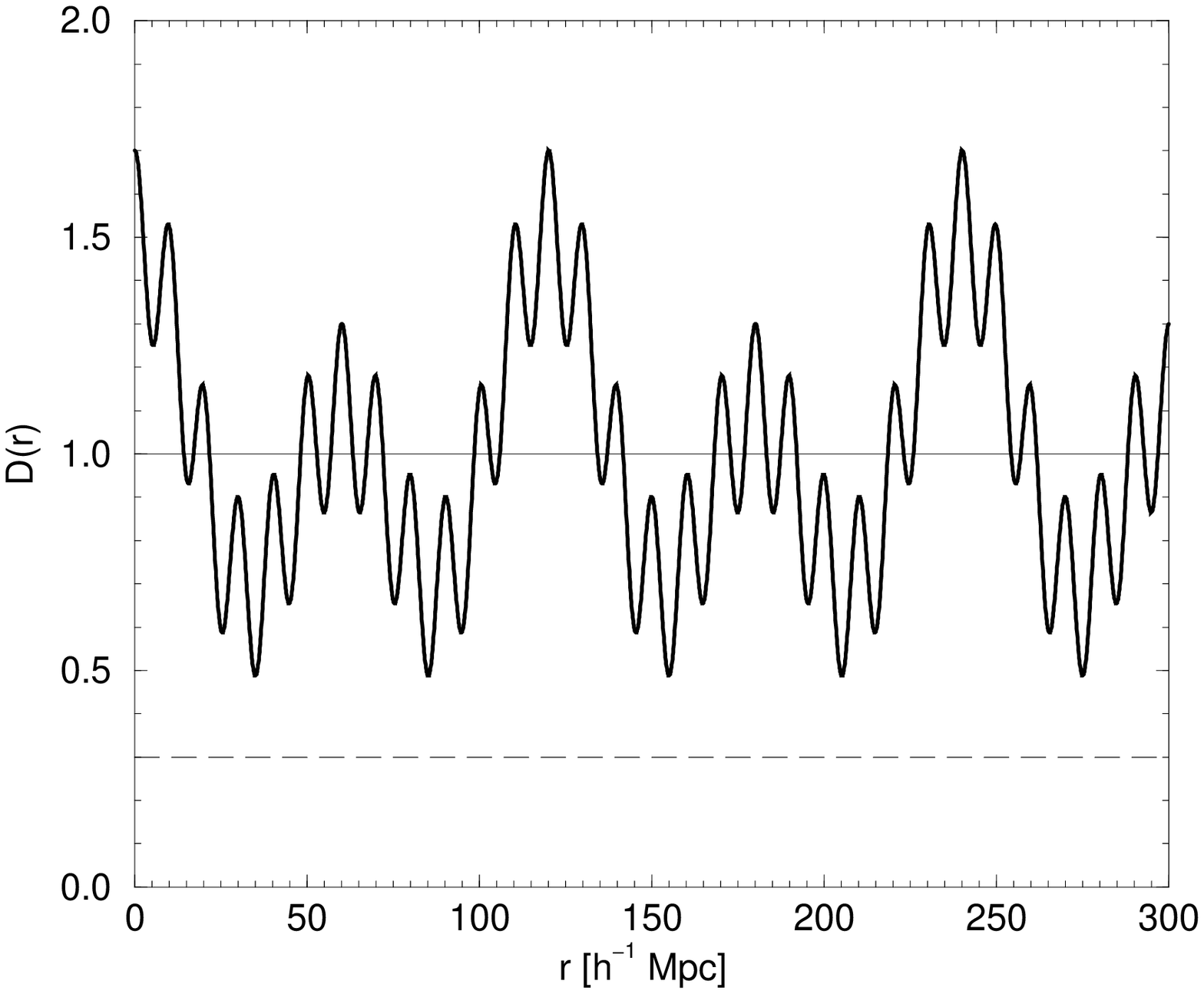}
\label{fig:saar}
\end{figure*}

Zeldovich \cite{z70} and Einasto, J\~oeveer \& Saar \cite{ejs80}
have shown that the density evolution of matter due to gravitational
instability is different in over-- and under--dense regions.  The
evolution follows approximately the law
\begin{equation}
D_c(t)={1 \over {1 - d_0 t/t_0}};
\label{eq:evol}
\end{equation}
where $d_0$ is a parameter depending on the amplitude of the density
fluctuations. In over--dense regions $d_0>0$, and the density
increases until the matter collapses and forms pancake or filamentary
systems \cite{bkp96} at a time $t_0$; thus the formula can be applied
only for $t \le t_0 $. In under--dense regions we have $d_0<0$ and the
density decreases, but never reaches zero (see Figure~\ref{fig:saar}).
In other words, there is always some dark matter in under--dense
regions.  At the time when over-dense regions collapse the density in
under--dense regions is half of the original (mean) density. In order
to form a galaxy the density of matter in a given region must exceed a
certain critical value \cite{ps74}, thus galaxies cannot form in
under--dense regions. They form only after the matter has flown to
over--dense regions: filaments, sheets, or clusters; here the
formation occurs {\em in situ}.

Consider the distribution of matter as a superposition of several
sinusoidal waves of amplitude $a_i$ and period $p_i$ around the mean
density $D_m$
\begin{equation}
D(r)=D_m + \sum_i a_i \sin(2\pi r/p_i).
\label{eq:sin} 
\end{equation}
Gravitational instability determines the evolution of these density
perturbations: large high over--dense regions become superclusters;
weakly over--dense regions become small filaments of galaxies and groups;
under--dense regions become voids, see Figure~\ref{fig:saar}. The fine
structure of superclusters is defined by perturbations of medium
wavelength, the structure of clusters by small--scale perturbations.

\section{Description functions}

Principal description functions that characterize the present
large--scale structure of the Universe are the power spectrum of
matter and galaxies, the correlation function of galaxies and
clusters, the cluster mass distribution, the void probability function
(VPF), and functions based on the clustering of galaxies and clusters
-- the multiplicity function, and the percolation function.  The
structure of the early Universe can be described by the angular
spectrum of the Cosmic Microwave Background (CMB). Here we discuss in
more detail properties of the power spectrum and correlation function,
and their dependence on geometrical properties of the distribution of
galaxies and clusters.

\subsection{Power spectrum}

The power spectrum describes the fluctuating density field
$\delta(x)$  through  its Fourier components $\delta_{k}$
\begin{equation}
P(k) = \langle\vert\delta_k\vert^2\rangle. 
\label{eq:sp}
\end{equation}
Here $k$ is the wavenumber in units of $h$~Mpc$^{-1}$.  The power
spectrum can be characterised by the power index on large scales, $n$,
and by its amplitude on certain characteristic scales. For the last
purpose usually very large scales ($\sim 1000$~\Mpc) are used, where
the amplitude is fixed by CMB observations by the COBE satellite
\cite{Bun97}, and scales where the power spectrum becomes non--linear,
$r \approx 8$~\Mpc.  The amplitude of the power spectrum on this scale
can be expressed through the $\sigma_8$ parameter -- which denotes the
rms density fluctuations within a sphere of radius 8~\Mpc.  It can be
calculated by integrating the power spectrum of matter.

The comparison of the distribution of real galaxies and particles in
simulations has shown that there are no luminous galaxies in voids;
here the matter has remained in its primordial dark form.  Now we
consider the influence of the presence of dark matter in voids on the
power spectrum of the clustered matter associated with galaxies.
Low-density matter in voids forms a smooth background of almost
constant density.  The density contrast of matter $\delta(x)$ can be
expressed as $\delta(x)=\left(D(x) - D_m \right)/D_m$, where $D(x)$ is
the density at location $x$, and $D_m$ is the mean density of matter
averaged over the whole space under study.  If we exclude from the
sample of all particles a population of approximately constant density
(void particles, see horizontal line in the right panel of
Figure~\ref{fig:saar}), but preserve all particles in high-density
regions, then the amplitudes of {\em absolute} density fluctuations
remain the same (as they are determined essentially by particles in
high-density regions), but the amplitudes of {\em relative}
fluctuations with respect to the mean density increase by a factor
which is determined by the ratio of mean densities, i.e. by the
fraction of matter in the new density field with respect to the
previous one
\begin{equation}
\delta(x)={D(x) - D_c \over D_c} {D_c \over D_m}, 
\label{eq:delta}
\end{equation}
here $D_c/D_m = F_c$ is the fraction of matter in the clustered
(galaxy) population.  A similar formula holds for the density contrast
in Fourier space, and we obtain the relation between power spectra of
matter and the clustered population
\begin{equation}
P_m(k) =  F_c^2 P_c(k).
\label{eq:sp2}
\end{equation}
We define the biasing parameter of galaxies (actually of all clustered
matter associated with galaxies) relative to matter through the ratio
of power spectra of galaxies and matter.  Both spectra are functions
of the wavenumber $k$, thus the biasing parameter is a function of
$k$.  Numerical simulations by Einasto \etal\  \cite{e99b} show that in the
linear regime of the structure evolution the biasing parameter is
practically constant for wavenumbers smaller than $k \approx
0.8$~\hmpc\ (scales larger than about 8~\Mpc).  Its value found from
simulations is very close to the expected value calculated from
Equation~(\ref{eq:sp2}). We come to the conclusion that the biasing
parameter is determined by the fraction of matter in the clustered
population
\begin{equation}
b_c=1/F_c.
\label{eq:bias}
\end{equation}

\begin{figure*}[ht]
\vspace*{5.7cm}
\caption{Left: power spectra of galaxies and clusters of galaxies
normalized to the amplitude of the 2-D APM galaxy power spectrum.  For
clarity error bars are not indicated and spectra are shown as smooth
curves rather than discrete data points.  Bold lines show spectra for
clusters data. Points with error bars show the spectrum of Abell
clusters by Miller \& Batuski \cite{mb00} adjusted to the galaxy
spectrum amplitude by a relative bias factor $b=3.2$.  Right:
correlation function of Abell clusters located in superclusters with
at least 8 clusters \cite{e97b}.  
}
\includegraphics{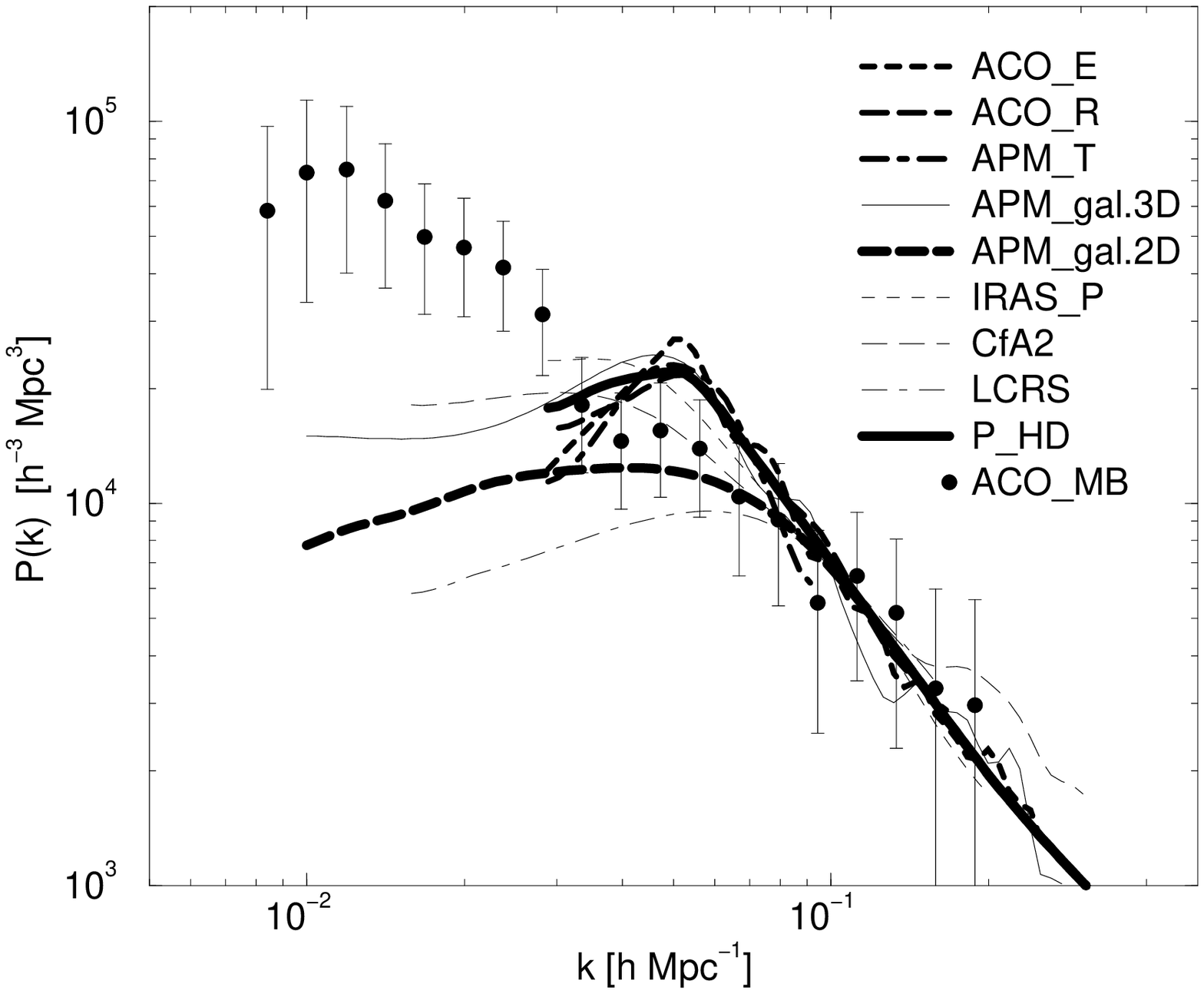} 
\includegraphics{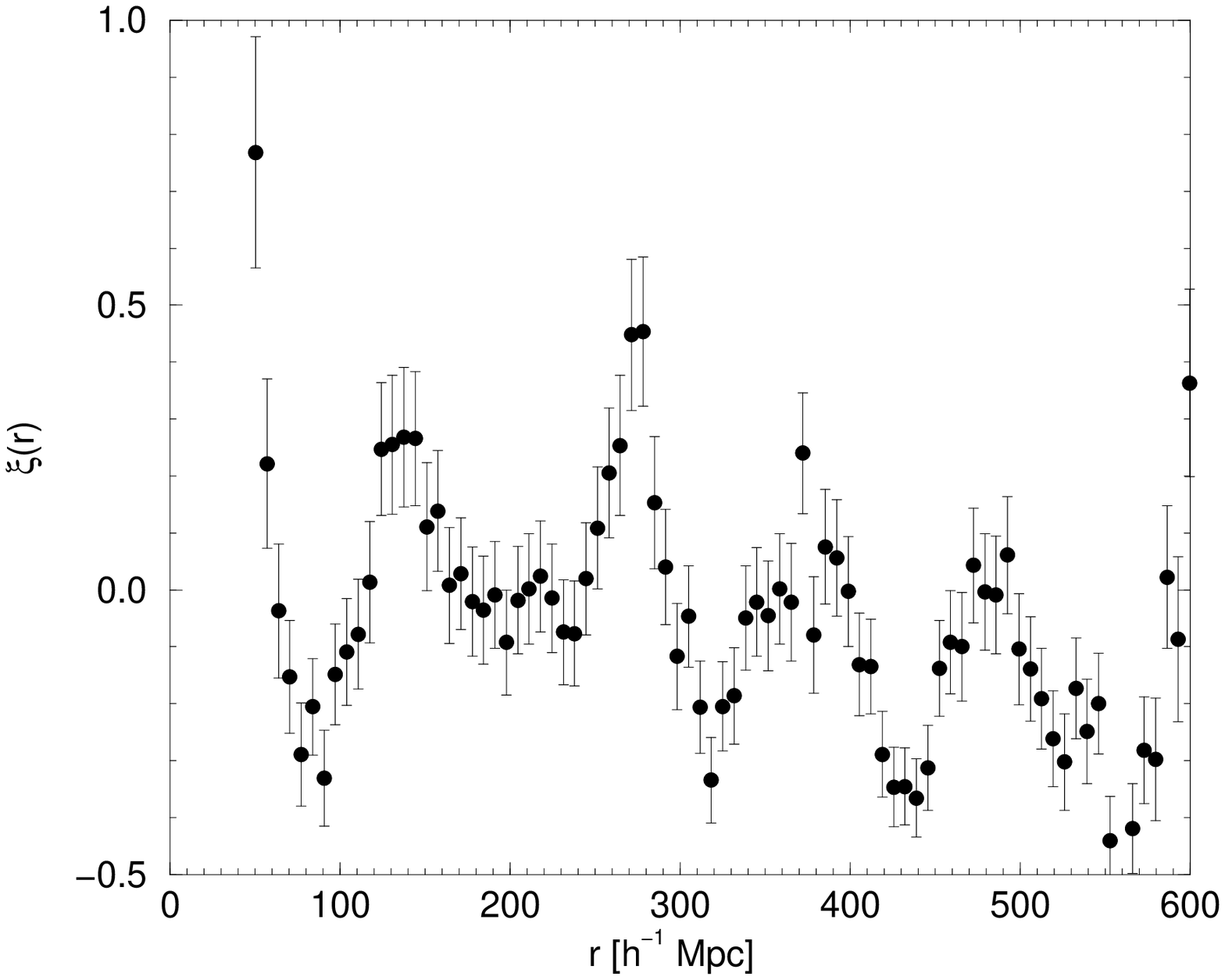}
\label{fig:sp_obs}
\end{figure*}

A summary of recent observational data on power spectra of galaxies
and clusters of galaxies of various type is shown in the left panel of
Figure~\ref{fig:sp_obs}, according to a compilation by Einasto \etal\
\cite{e99a}.  Galaxy and cluster spectra are adjusted in amplitude to
reduce them to the power spectrum of APM galaxies.  We assume that the
amplitude of the power spectrum of APM galaxies represents well the
amplitude of the whole clustered (galaxy) population.  On large scales
we use a recent determination \cite{mb00} on the basis of Abell
clusters of richness class 1 and higher.

\subsection{Correlation function}

The two-point correlation function $\xi(r)$ is defined as the excess
over Poisson of the joint probability of finding objects in two volume
elements separated by $r$ and averaged over a very large volume
\cite{p80}.  We shall use the term ``correlation function'' for
its estimate, determined in a limited volume, and calculate it using
the formula:
\begin{equation}
\xi(r)={\langle DD(r)\rangle  \over \langle RR(r)\rangle }{n_R^2
\over n^2} -1, 
\label{eq:corr}
\end{equation}
where $\langle DD(r)\rangle $ is the number of pairs of galaxies (or
clusters of galaxies) in the range of distances $r\pm dr/2$, $dr$ is
the bin size, $\langle RR(r)\rangle $ is the respective number of
pairs in a Poisson sample of points, $n$ and $n_R$ are the mean number
densities of clusters in respective samples, and brackets $\langle
\dots \rangle $ denote the ensemble average. The summation is over the
whole volume under study, and it is assumed that the galaxy and
Poisson samples have identical shape, volume and selection function.

Both the correlation function and the power spectrum characterise the
distribution of galaxies, clusters and superclusters.  In an ideal
case in the absence of errors, and if both functions are determined in
the whole space, they form a mutual pair of Fourier transformations:
\begin{equation}
\xi(r)={1 \over 2\pi^2}\int_0^{\infty}{ P(k) k^2 {\sin kr \over kr} dk},
\label{eq:cf2}
\end{equation}
\begin{equation}
P(k)=4\pi\int_0^{\infty}{\xi(r) r^2 {\sin kr \over kr} dr}.
\label{eq:sp3}
\end{equation}
These formulae are useful when studying theoretical models or results
of numerical simulations.  For real samples they are of less use since
observational errors and selection effects influence these functions
in a different way.  Also they reflect the spatial distribution of
objects differently, thus they complement each other.

There exists a large body of studies of the correlation function of
galaxies and clusters.  Already early studies have shown that on small
scales the correlation function can be expressed as a power law  (see
\cite{p80}):
\begin{equation}
\xi(r)=(r/r_0)^{-\gamma},
\label{eq:cf3}
\end{equation}
where $\gamma \approx 1.8$ is the power index, and $r_0 \approx
5$~\Mpc\ is the correlation length.  The correlation function of
clusters of galaxies is similar, but shifted to larger scales, i.e. it
has approximately the same power index but a larger correlation length,
$r_0 \approx 25$~\Mpc.  On small scales the correlation function
reflects the fractal dimension, $D=3-\gamma$, of the distribution of
galaxies and clusters \cite{ss85}.  On  large scales the
correlation function depends on the distribution of systems of
galaxies.

In order to understand better how different geometries of the
distribution of galaxies and clusters are reflected in the properties of
the correlation function and power spectrum, we shall construct
several mock samples with known geometrical properties, and calculate
both functions. Here we  use results of the study of geometrical
properties of the correlation function \cite{e91,e97b}.

\begin{figure}[ht]
\vspace*{17.5cm} 
\caption{The distribution of galaxies and clusters in random mock models
(upper left and right panels, respectively), respective correlation
functions (middle panels) and power spectra (lower panels). For
explanations see text.}
\includegraphics{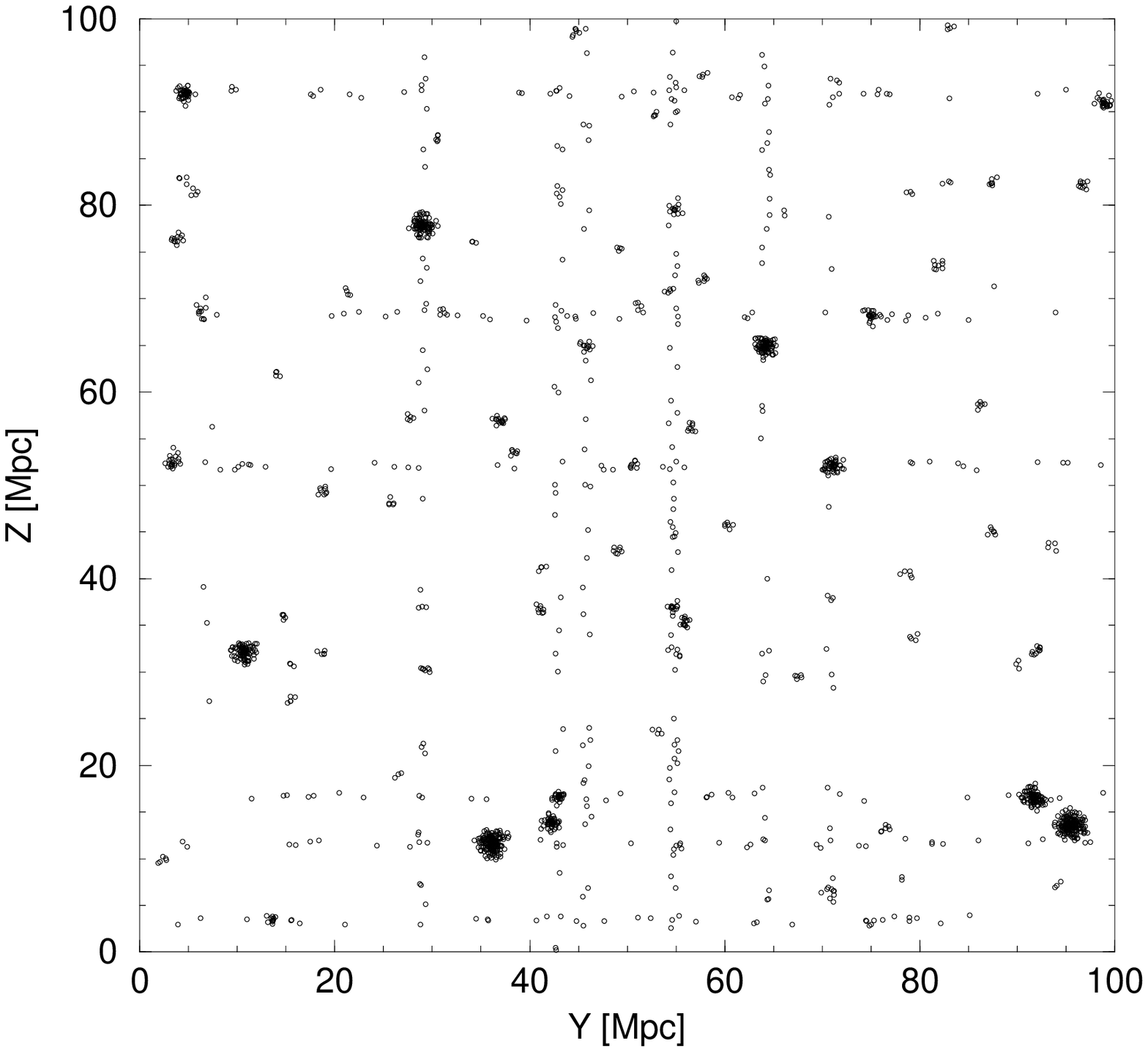} 
\includegraphics{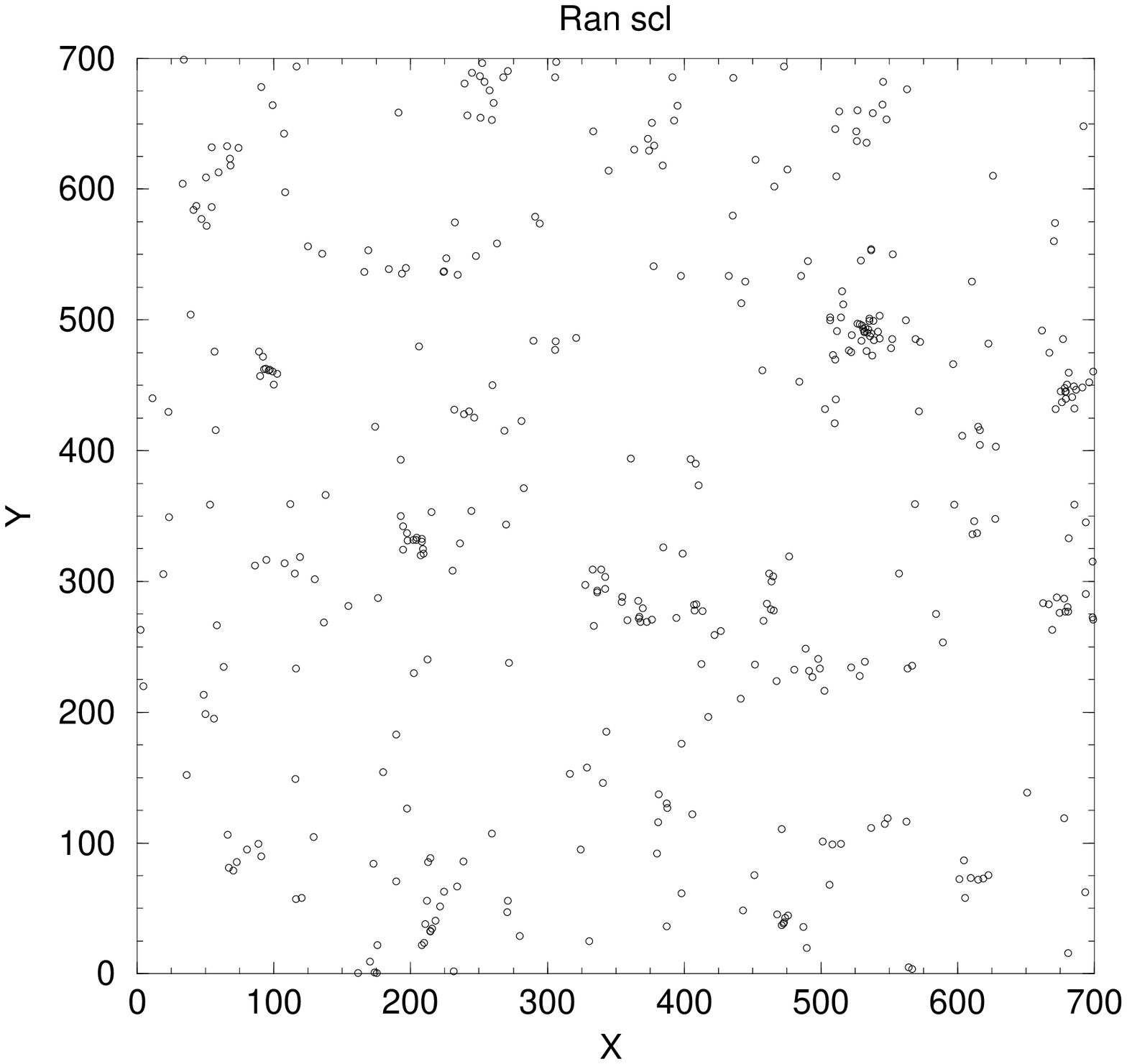}
\includegraphics{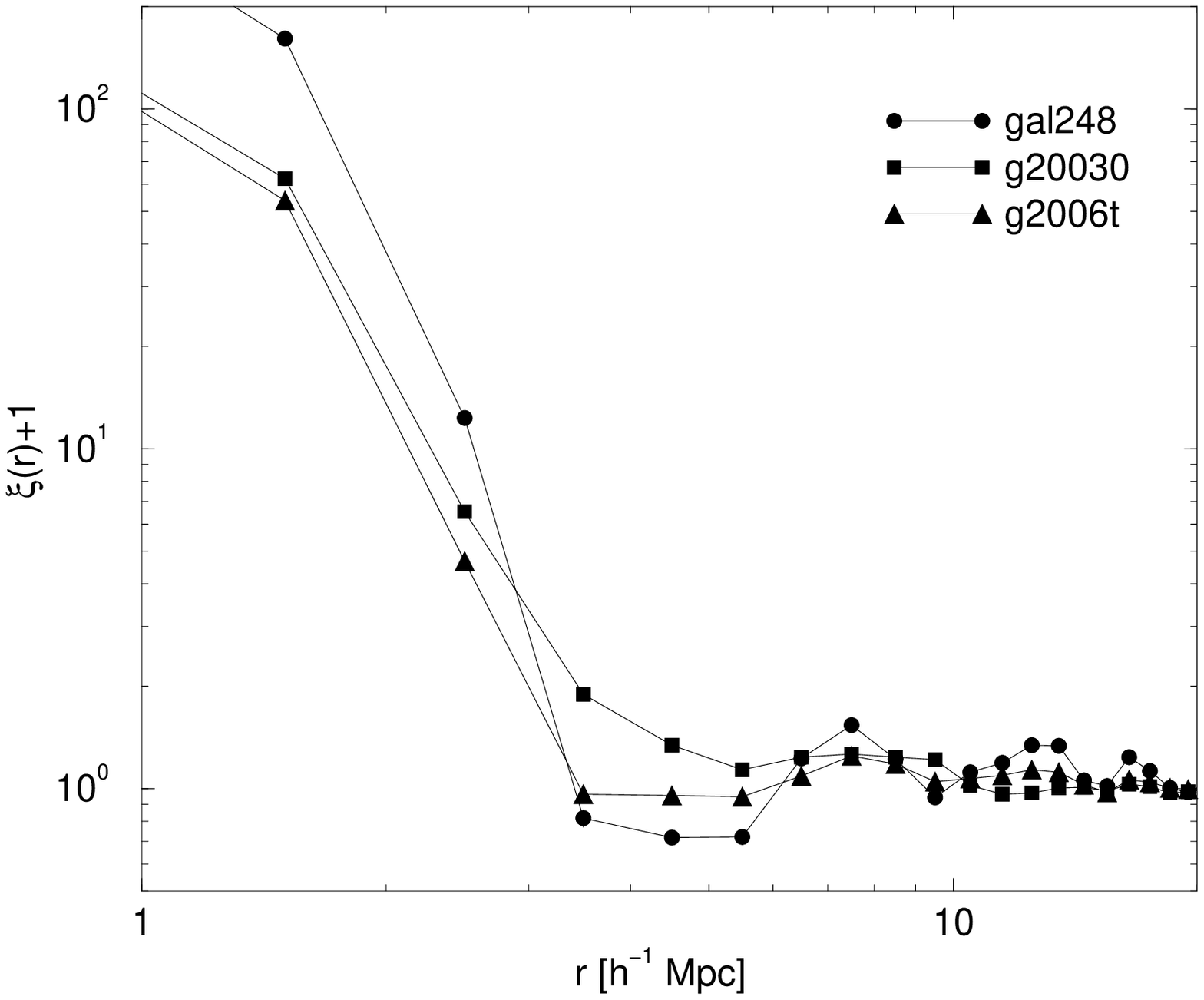} 
\includegraphics{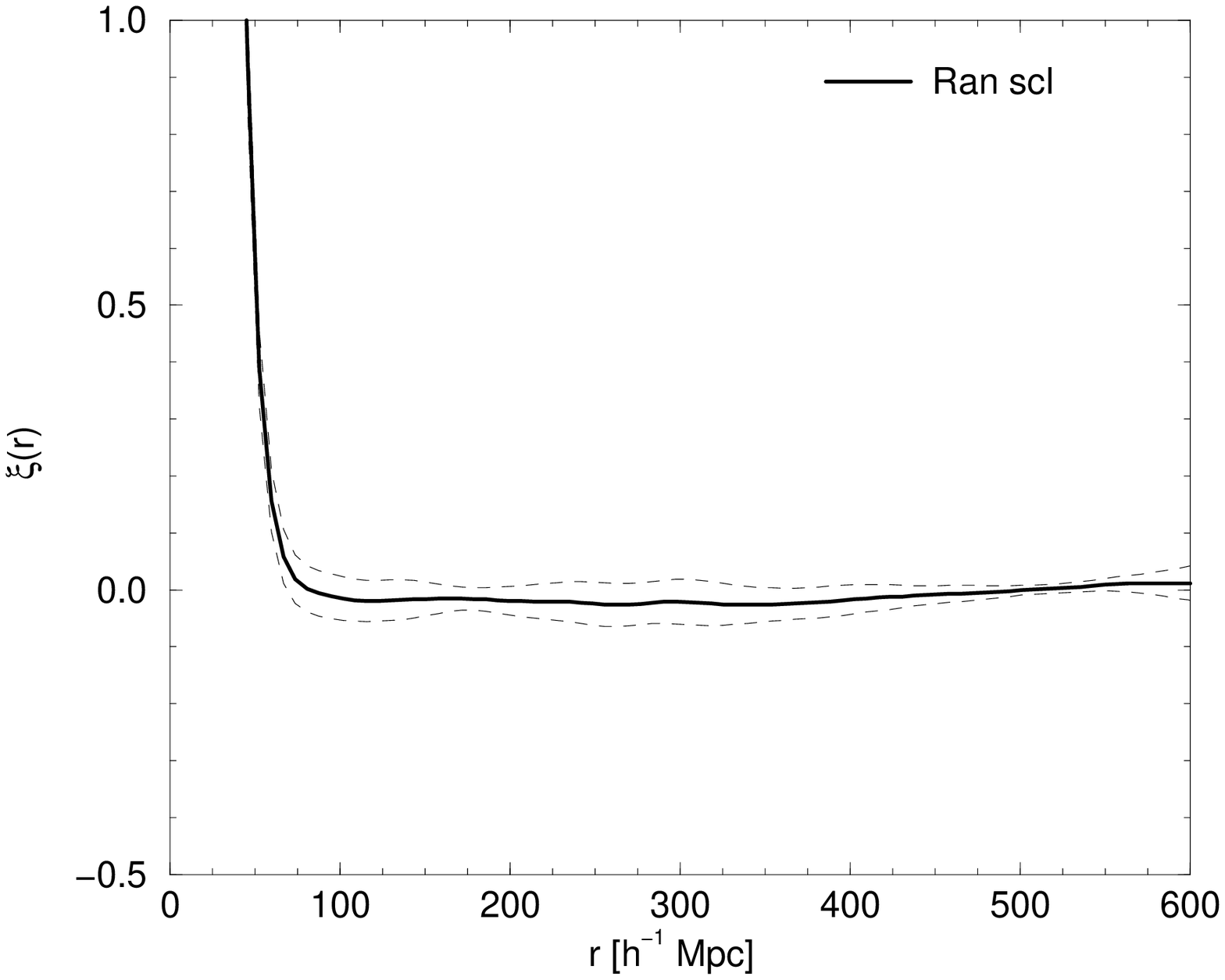}
\includegraphics{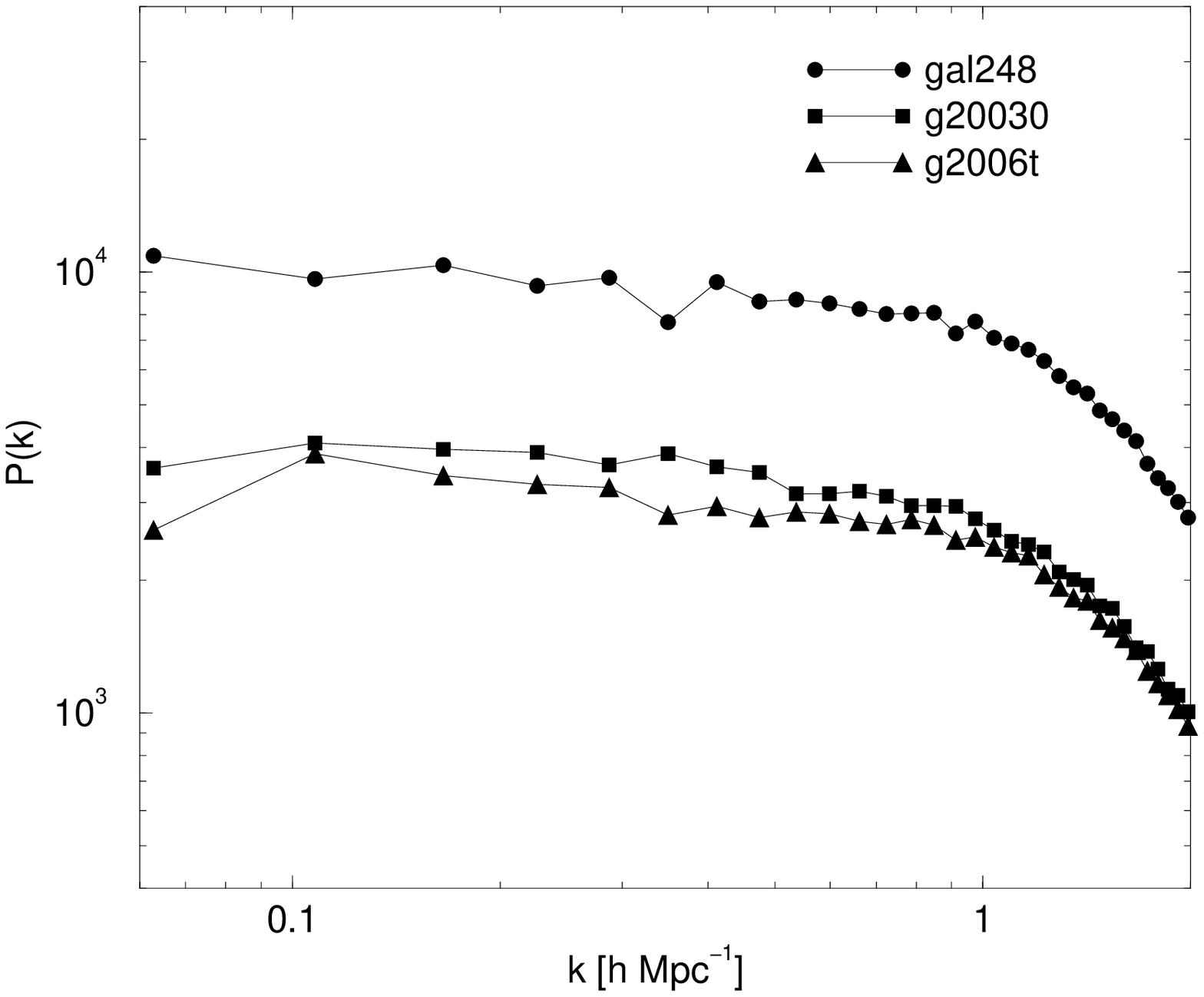} 
\includegraphics{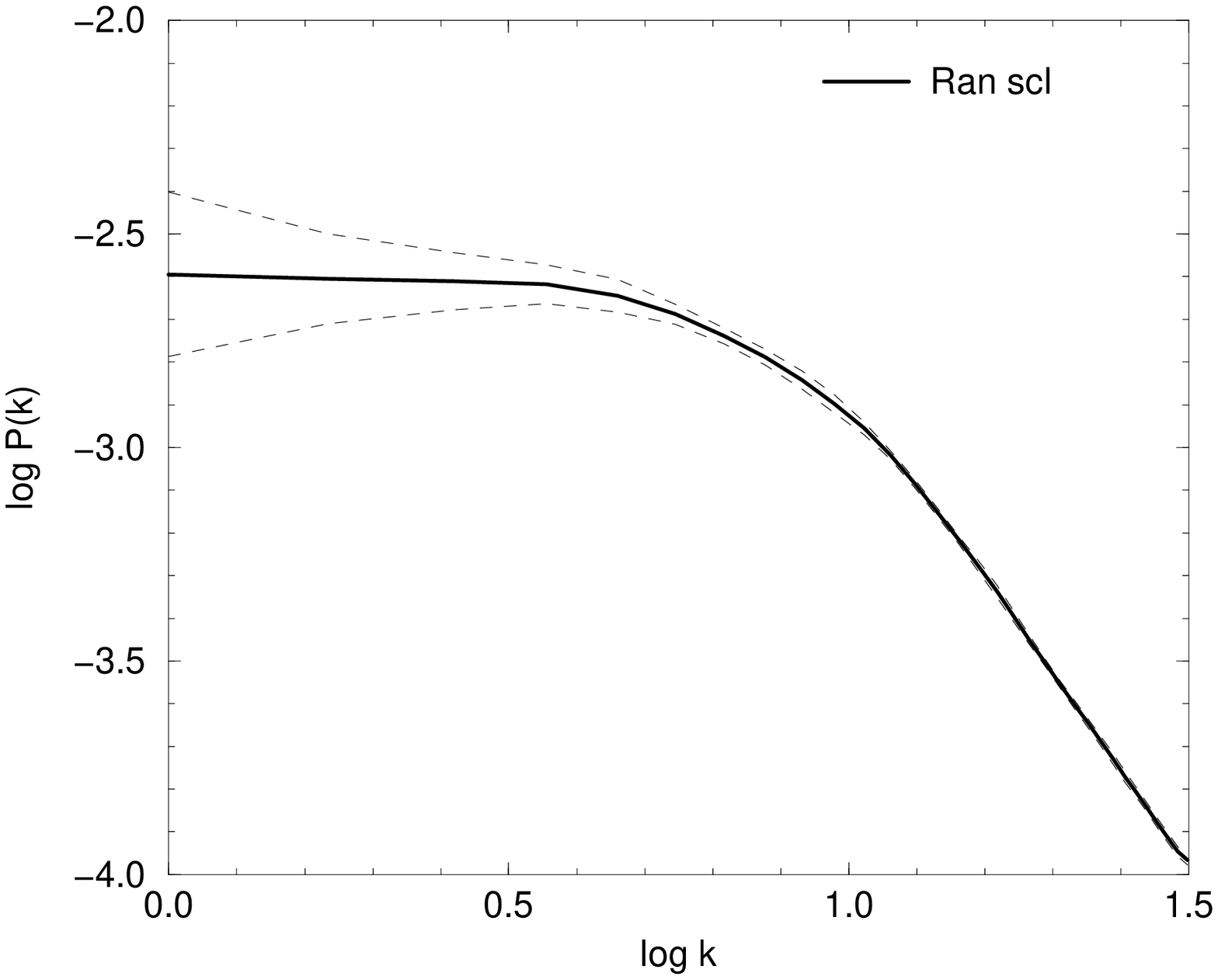}
\label{fig:corr1}
\end{figure}

\begin{figure}[ht]
\vspace*{17.5cm} 
\caption{The distribution of  clusters in Voronoi and regular mock models
(upper left and right panels, respectively), respective correlation
functions (middle panels) and power spectra (lower panels). For
explanations see text.}
\includegraphics{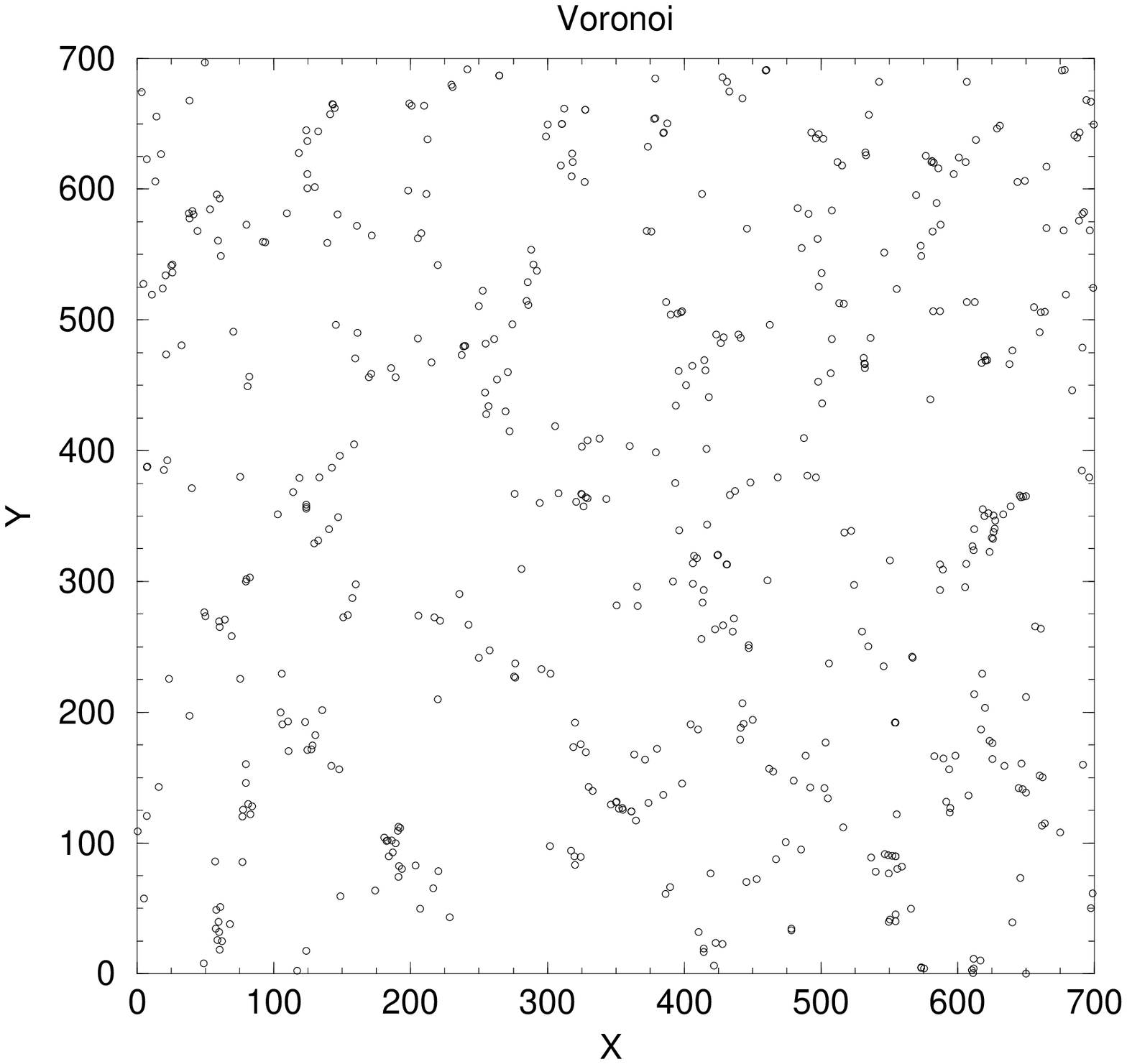} 
\includegraphics{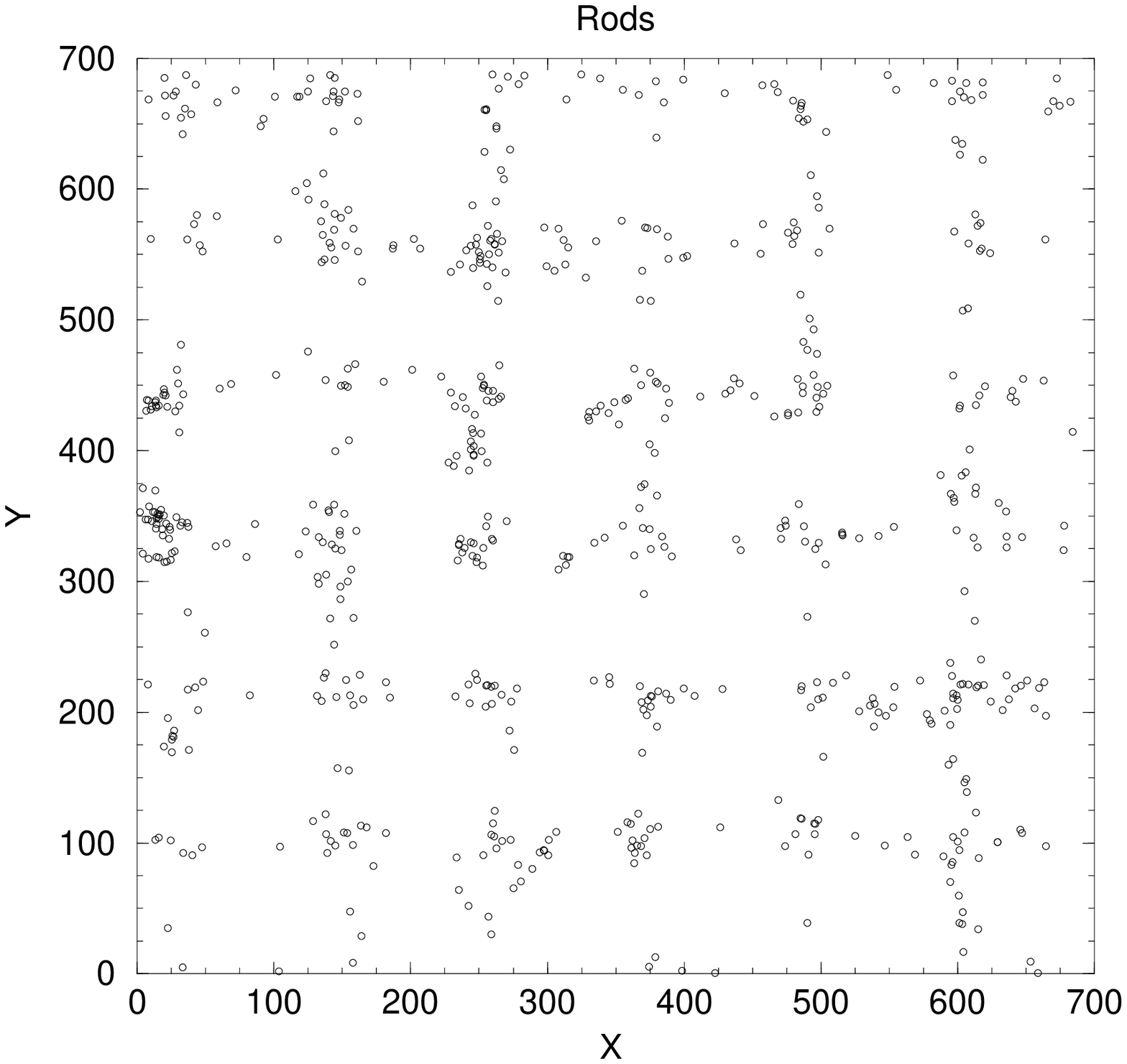}
\includegraphics{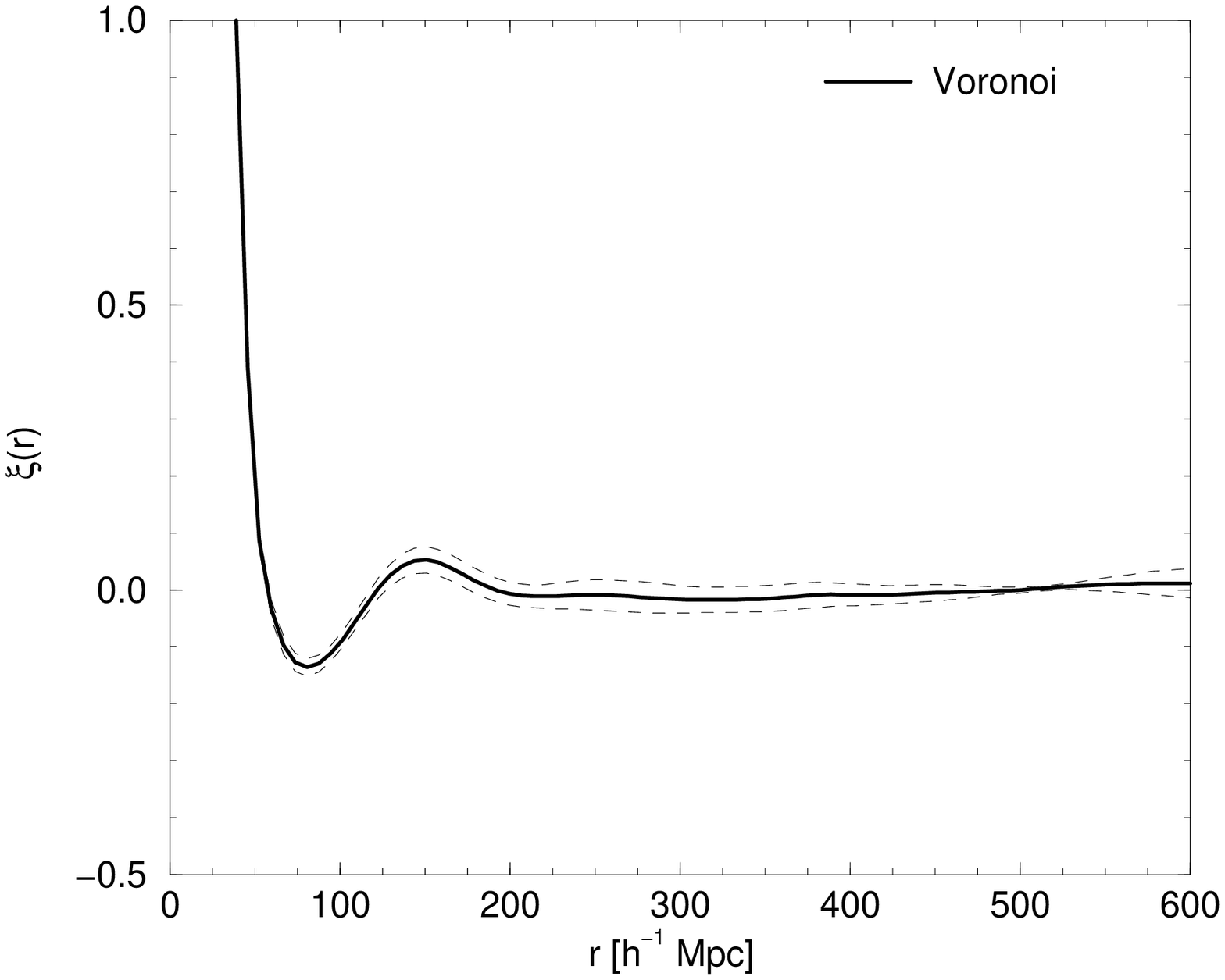} 
\includegraphics{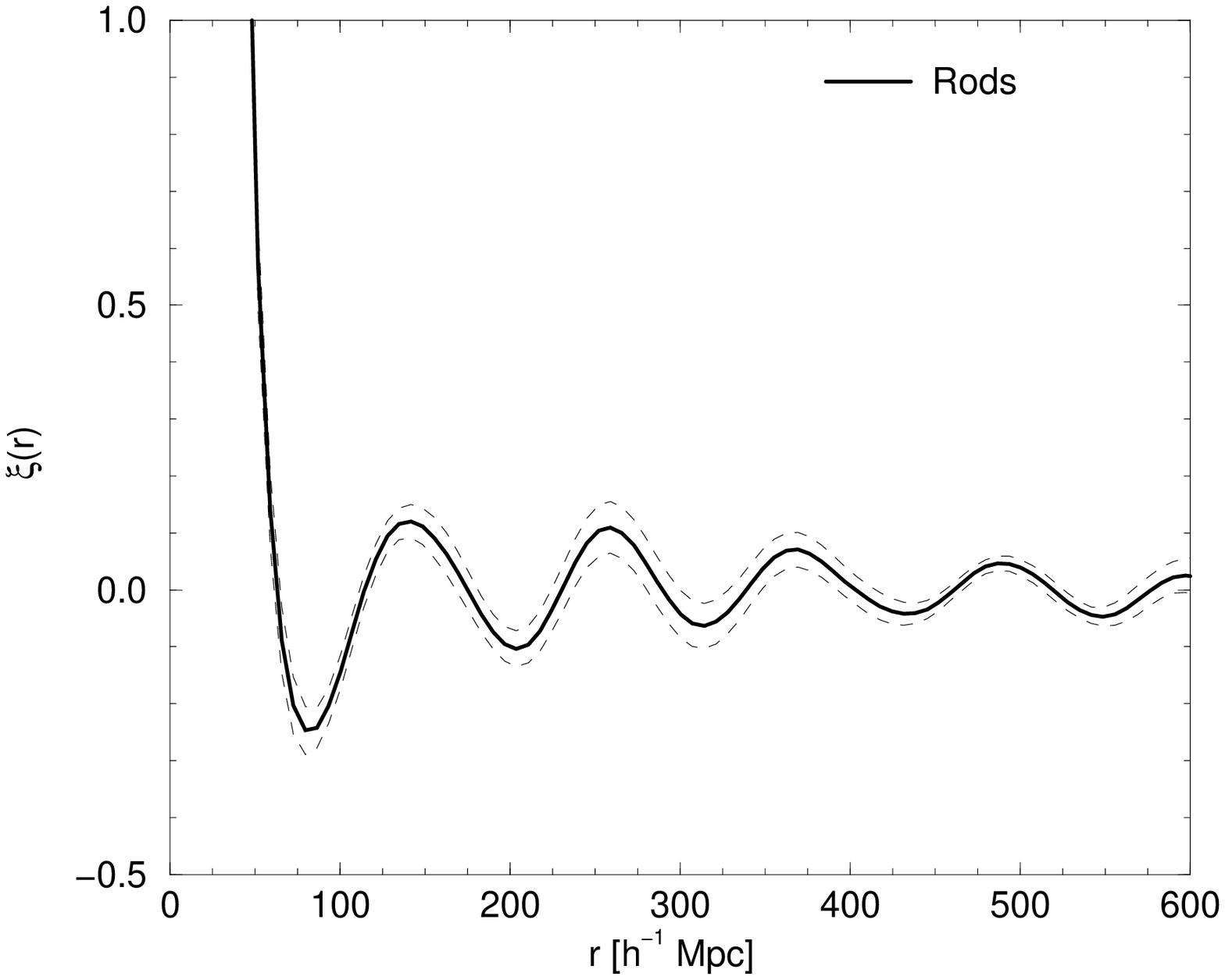}
\includegraphics{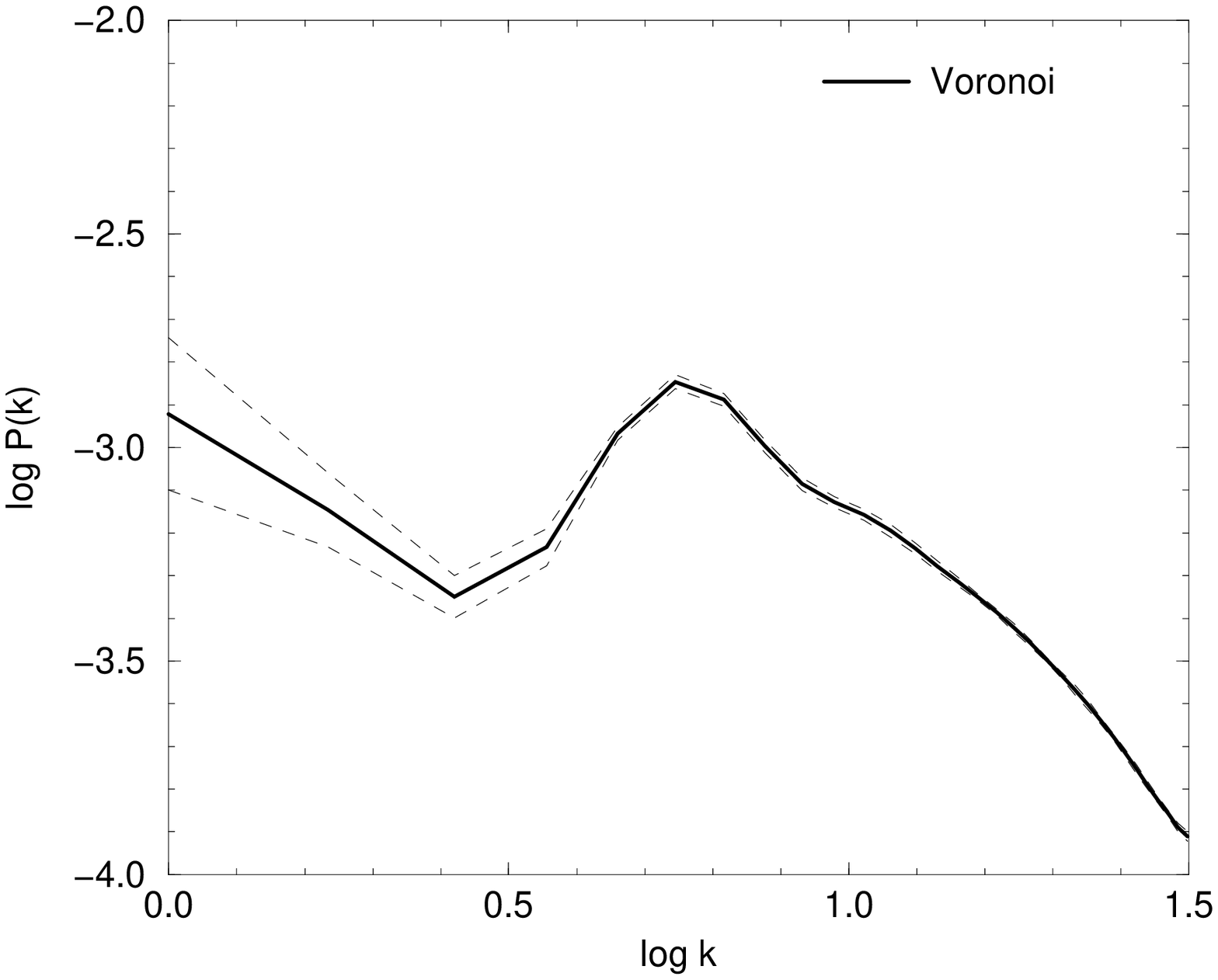} 
\includegraphics{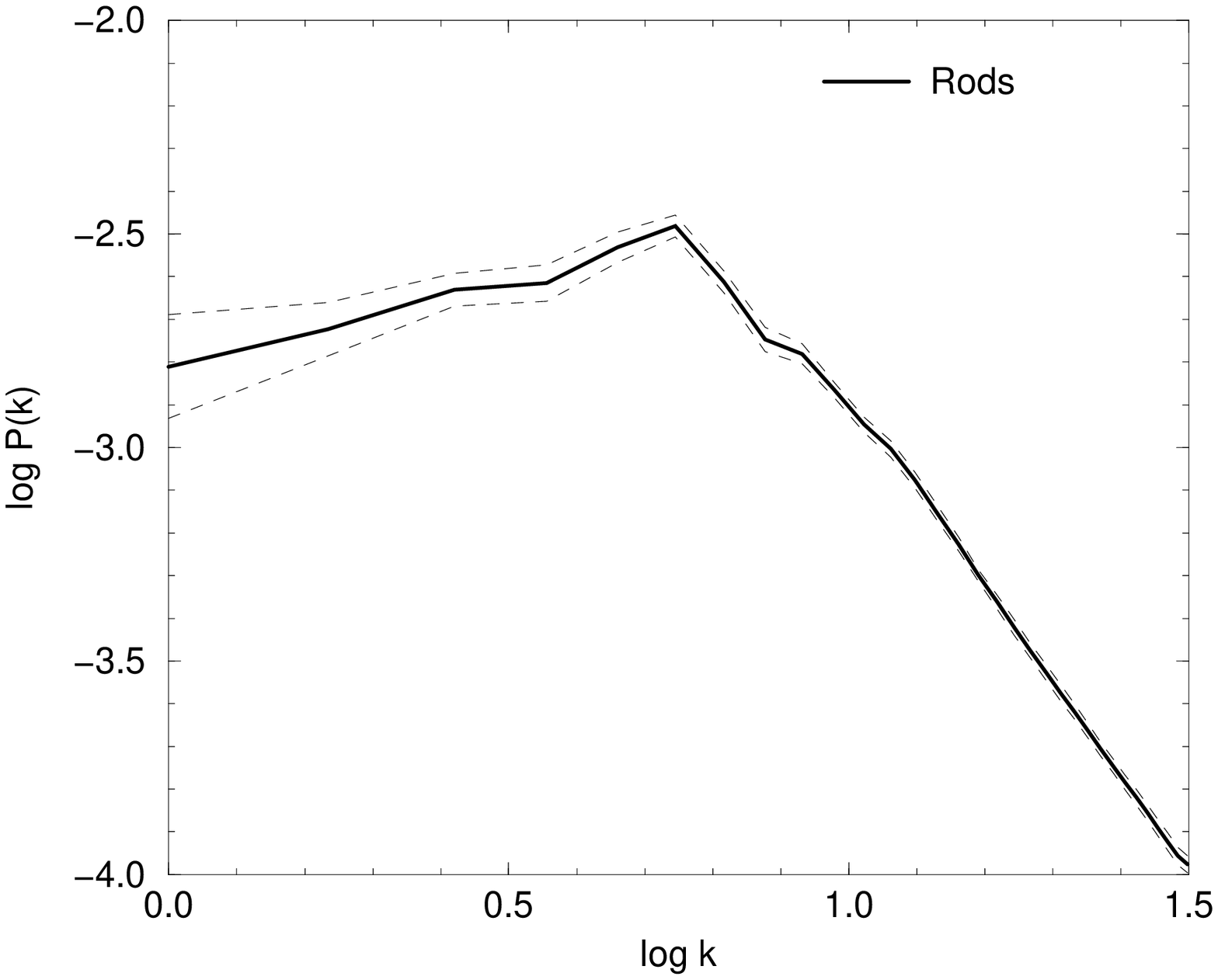}
\label{fig:corr2}
\end{figure}

The correlation function is determined by mutual distances of galaxies
(or clusters) in real space, thus this function depends directly
on the structure of galaxy systems themselves (on small separations
which are comparable to sizes of these systems), and on the
distribution of galaxy systems (on separations which exceed the
dimensions of galaxy systems).  To see these effects separately we
have constructed two series of mock samples.  In the first series we
change the distribution of galaxies on small scales following
\cite{e91}, in the second series we change the distribution of systems
themselves \cite{e97b}.  In the first case we consider test particles
as galaxies; they form two populations -- clusters and field galaxies.
Clusters are located randomly in a cube of size $L=100$~\Mpc; clusters
have a varying number of member galaxies from 12 to 200, and an
abundance and mass distribution in accordance with the observed
cluster mass distribution (see the next subsection).  Inside clusters
galaxies are located randomly with an isothermal density distribution.
For the field population we consider three cases.  In the first model
there is no field population at all; in Figure~\ref{fig:corr1} this
model is designated gal248 (it contains 248 clusters). The second
model has 200 clusters and 6000 randomly located field galaxies
(designated as g2006t). The third model has also 200 clusters, but
field galaxies are located in filaments -- each filament crosses one
cluster in either x, y, or z--axis direction, randomly chosen, and has
30 galaxies; this sample is designated as g20030.  The total number of
galaxies in clusters and in the field population in the two last
models is approximately equal.

The distribution of galaxies in a sheet of thickness 15~\Mpc\ of the
model g20030 is shown in the upper left panel of
Figure~\ref{fig:corr1}.  In the other two models the distribution of
clusters is similar, but field galaxies are distributed randomly or
are absent.  The correlation functions and power spectra of all three
models are plotted in the middle and lower left panels, respectively.
We see that the slope (power law index) of the correlation function at
small separations is $\gamma\approx 3$ in the pure cluster model, and
$\gamma\approx 1.8$ in other two models.  These differences are due to
the difference in the effective fractal dimension: $D=3 -\gamma
\approx 0$ for the pure cluster model (clusters are spherically
symmetrical); and $D \approx 1.2$ in a mixture of spherically
symmetrical clusters and one--dimensional filaments.  The power index
depends on the partition of galaxies between the clustered and the
field populations.  On scales of $r \approx 3$~\Mpc\ the correlation
function changes sharply. In the pure cluster model it has small
negative values for $r > 3$~\Mpc, while in the random field models
$\xi(r) \approx 0$ for $r > 3$~\Mpc. In the model with filaments the
correlation function is positive but has a smaller power index, of
$\gamma \approx 1.2$.  The change of the power index occurs on a scale
equal to the diameter of clusters of galaxies.  On larger scale the
behaviour of the correlation function depends on properties of the
field population and on the distribution of clusters.  In the absence
of the field there are only a few close neighbours of clusters, hence
the slightly negative value of the correlation function; on larger
scales the level of the correlation function reflects the distribution
of clusters; clusters are distributed randomly and the level is
approximately zero.  In the random field model galaxies are located
also in the vicinity of clusters and the zero level of the correlation
function begins immediately beyond the end of cluster galaxies.  In
the filamentary model the effective power index of the correlation
function reflects the mean fractal dimension of filaments.
   
The power spectrum of all three models has a similar shape.  On small
scales the power index is negative and depends on the clustering law
of galaxies in clusters.  On larger scales the shape of the power
spectrum depends on the distribution of galaxies and clusters on
respective scales.  Because both clusters and field galaxies are
essentially randomly distributed (the location of filaments is also
random), the spectrum has a zero power index as expected for a random
distribution.  The amplitude of the power spectrum depends on the
fraction of galaxies in clusters.  In the pure cluster model the
amplitude is much higher, and the difference in the amplitude depends on
the fraction of galaxies in clusters.  This effect is similar to the
influence of the void population discussed above.

Now we shall discuss models of the cluster distribution.  We do not
include the population of galaxies; instead we shall investigate how
different distribution of clusters affects the correlation function
and power spectrum.  Here we assume that a fraction of clusters are
located in superclusters, the rest form a field cluster population.
We consider again three models: randomly distributed superclusters,
regularly spaced superclusters, and superclusters formed by the
Voronoi tessellation model.  In the Voronoi model centers of voids are
located randomly, and clusters are placed as far from void centres as
possible.  These models differ in their degree of regularity of
the distribution of superclusters.  The random supercluster model has
no regularity and no built--in scale.  The Voronoi model has a
characteristic scale -- the mean diameter of voids (determined by the
number of voids in the sample volume), but no regularity in the
distribution of voids.  In the regular model superclusters are located
randomly along rods which form a regular rectangular grid of step size
120~\Mpc; this scale defines the mean size of voids between
superclusters, and also puts voids to a semiregular honeycomb--like
lattice.  In addition a field population of isolated randomly located
clusters is present in this model.

The distribution of clusters of the random, Voronoi, and regular model
are shown in the upper panels of Figure~\ref{fig:corr1} and
~\ref{fig:corr2}; correlation functions are given in the middle panels, and
power spectra in the lower panels.  We see that on small scales all
correlation functions are identical; power spectra on small
wavenumbers are also identical.  This is due to the fact that on these
scales both functions are determined by the distribution of clusters
{\em within} superclusters; and superclusters in all models were
generated using the same algorithm as in generating galaxies in
clusters.  On larger scales there are important differences between
models.  In the random supercluster model the correlation function
approaches zero at $r > 80$~\Mpc. In the Voronoi model it has a
minimum around $r \approx 80$~\Mpc, followed by a secondary maximum at
$r \approx 150$~\Mpc; thereafter it approaches zero.  The correlation
function of the regular rod model is oscillating: it has a series of
regularly spaced maxima and minima with a period of $\sim 120$~\Mpc;
the amplitude of oscillations decreases with increasing separation.
The power spectrum of the random supercluster model is flat and
featureless on large scales, while in the Voronoi and regular models it has
a sharp maximum at wavenumber, which corresponds to the mean diameter
of voids in models and to the period of oscillations of the
correlation function.  The shape of the power spectrum on large scales
of these two models is, however, different.

These mock samples illustrate properties of the correlation function
and power spectrum on small and large scales and their dependence on
the distribution of galaxies and clusters within systems and on the
distribution of systems themselves.  The correlation function of
clusters of galaxies in rich superclusters is shown in the right panel
of Figure~\ref{fig:sp_obs}.  This function is oscillating with a
period which corresponds to the maximum of the power spectrum at
$k=0.05$~\hmpc, seen in the left panel of the same Figure.  We note
that a periodicity of the distribution of high--density regions with
the same period has been observed in the direction of the galactic poles
by Broadhurst \etal\  \cite{beks}.  All these facts suggests that there
exists a preferred scale of $\sim 130$~\Mpc\ in the Universe, and
possibly also some regularity in the distribution of the
supercluster--void network.

\subsection{Mass function of clusters of galaxies}

Masses of clusters of galaxies can be determined from the velocity
dispersion of its member galaxies, or on the basis of their X--ray
emission (using the hot gas as an indicator of the velocity
dispersion in clusters), or else from the gravitational lens effect.
Using masses of clusters and their abundance it is
possible to calculate the number of clusters of different mass,
$N(>M)$.  This mass function is usually expressed in units of $h^{-1}
M_0$ in a sphere of radius 1.5~\Mpc.  The cluster mass function was
derived by Bahcall \& Cen \cite{bc93}, and also by Girardi \etal\
\cite{gir98}.  The function characterizes the distribution of systems
of galaxies at the present epoch.  There exist estimates of the
abundance of clusters at high redshifts, but they are still very
uncertain.  The function $N(>M)$, and its specific value,
$N(>10^{14}M_0)$, can be used to constrain cosmological parameters.
We shall use this constraint in the next Section.

\section{Cosmological parameters}

As cosmological parameters we consider parameters which define the
present and past structure of the Universe.  Principal parameters are:
the Hubble constant, which characterises the expansion speed of the
Universe; the age and acceleration parameter of the Universe;
densities of main constituents of the Universe: baryonic matter, dark
matter and dark energy; and parameters, which define the amplitude and
shape of the power spectrum of galaxies and matter.  Cosmological
parameters and descriptive functions can be used to test various
scenarios of structure evolution.

The Hubble constant, $h$, can be estimated by several methods: through
the ladder of various distance estimators from star clusters to
cepheids in nearby galaxies, through the light curves of
medium-distant supernovae, or using several physical methods
(gravitational lensing, Sunyaev--Zeldovich--effect).  Summaries of
recent determinations are given in \cite{petal00,setal00}.  A mean
value of recent determinations is $h= 0.65 \pm 0.07$.

The baryon density can be determined most accurately from observations
of the deuterium, helium and lithium abundances in combination with
the nucleosynthesis constrains.  The best available result is
$\Omega_b h^2 = 0.019 \pm 0.002$ \cite{b99}.

The total density of matter, $\Omega_{tot} = \Omega_m + \Omega_v$,
determines the position of the first Doppler peak of the angular
spectrum of CMB temperature fluctuations; here $\Omega_m$ and
$\Omega_v$ are the densities of matter and dark (vacuum) energy,
respectively. Recent observations show that the maximum of the first
Doppler peak lies at $l \approx 200$ \cite{deb00,h00}.  This
indicates that $\Omega_{tot} \approx 1$.  Since this is the
theoretically preferred value, I assume in the following that
$\Omega_{tot} = 1$.

There exist a number of methods to estimate the density of matter,
$\Omega_m = \Omega_b + \Omega_c + \Omega_n$, where $\Omega_b$,
$\Omega_c$, and $\Omega_n$ are the densities of baryonic matter, cold
dark matter (CDM), and hot dark matter (HDM), respectively.  The
luminosity--distance method, used in the distant supernova project,
yields $\Omega_m = 0.28 \pm 0.05$ \cite{petal98,retal98}.  Another
method is based on X-ray data on clusters of galaxies, which gives the
fraction of gas in clusters, $f_{gas} = \Omega_b/\Omega_m$.  If
compared to the density of the baryonic matter one gets the estimate
of the total density, $\Omega_m = 0.31 \pm 0.05 (h/0.65)^{-1/3}$
\cite{metal00}.  A third method is based on the geometry of the
Universe.  Observations show the presence of a dominant scale, $l_0 =
130 \pm 10$~\Mpc, in the distribution of high-density regions
\cite{beks,e97a,e97b}.  A similar phenomenon is observed in the
distribution of Lyman-break galaxies \cite{bj00} at high redshift, $z
\approx 3$.  We can assume that this scale is primordial and co-moves
with the expansion; in other words -- it can be used as a standard
ruler.  The relation between redshift difference and linear comoving
separation depends on the density parameter of the Universe; for a
closed universe one gets a density estimate $\Omega_m = 0.4 \pm 0.1$.
The same method was applied for the distribution of quasars by Roukema
\& Mamon \cite{rm00} with the result $\Omega_m = 0.3 \pm 0.1$.
Finally, the evolution of the cluster abundance with time also depends
on the density parameter (see \cite{bops} for a review). This method
yields an estimate $\Omega_m = 0.4 \pm 0.1$ for the matter density.
The formal weighted mean of these independent estimates is $\Omega_m =
0.32 \pm 0.03$.

Cosmological parameters enter as arguments in a number of functions
which can be determined from observations.  These functions include
the power spectrum of galaxies, the angular spectrum of temperature
fluctuations of the CMB radiation, the cluster mass and velocity
distributions.  I accept the power spectrum of galaxies according to a
summary in \cite{e99a} with the addition of the recent determination
of the cluster power spectrum \cite{mb00}.  The amplitude of the power
spectrum can be expressed through the $\sigma_8$ parameter -- rms
density fluctuations within a sphere of radius 8~\Mpc. This parameter
was determined for the present epoch for galaxies, $(\sigma_8)_{gal} =
0.89 \pm 0.09$ \cite{e99a}.  For the CMB angular spectrum I use recent
BOOMERANG and MAXIMA I measurements \cite{deb00,h00}.  For the cluster
mass distribution I use the determinations by \cite{bc93} and
\cite{gir98}.

\section{Cosmological models}

The power spectra of matter and the angular spectra of CMB can be
calculated for a set of cosmological parameters using the CMBFAST
algorithm \cite{sz96}; spectra are COBE normalized.  The cluster
abundance and mass distribution functions can be calculated by the
Press-Schechter \cite{ps74} algorithm.  We have used these
algorithms to test how well cosmological parameters are in agreement
with these descriptive functions.

One problem in comparing cosmological models with observations is
related to the fact that from observations we can determine the power
spectra and correlation functions of galaxies and clusters of
galaxies, but using models we can do that for the whole matter.  
Power spectra of galaxies and matter are related
through the bias parameter.  There exist various methods to estimate
the bias parameter, using velocity data.  Here we use another method
which is based on the numerical simulation of the evolution of the
Universe.  During dynamical evolution matter flows away from
low-density regions and forms filaments and clusters of galaxies.
This flow depends slightly on the density parameter of the model. The
fraction of matter in the clustered population can be found by
counting particles with local density values exceeding a certain
threshold.  To separate void particles from clustered particles we
have used the mean density, since this density value divides regions
of different cosmological evolution, see eq.~(\ref{eq:evol}).
Hydrodynamical simulations by Cen \& Ostriker \cite{co92} confirmed
that galaxy formation occurs only in over--dense regions.

\begin{figure}[ht]
\vspace*{11.5cm}
\caption{ Upper left: the fraction of matter in the clustered population
associated with galaxies as a function of $\sigma_8$ for 2 LCDM models
(dashed curves); and the relation between $F_{gal}$ and
$(\sigma_8)_m$ (bold solid line). Upper right: the biasing parameter
needed to bring the amplitude $\sigma_8$ of the model into
agreement with the observed $\sigma_8$ for galaxies and for LCDM and MDM
models with various matter density $\Omega_m$ and HDM density,
$\Omega_n$. The dashed box shows the range of the bias parameter allowed by
numerical simulations of the evacuation of voids. Lower left: power
spectra of LCDM models with various $\Omega_m$. Lower right: angular
spectra of CMB for LCDM and MDM models for various $\Omega_m$.  }
\includegraphics{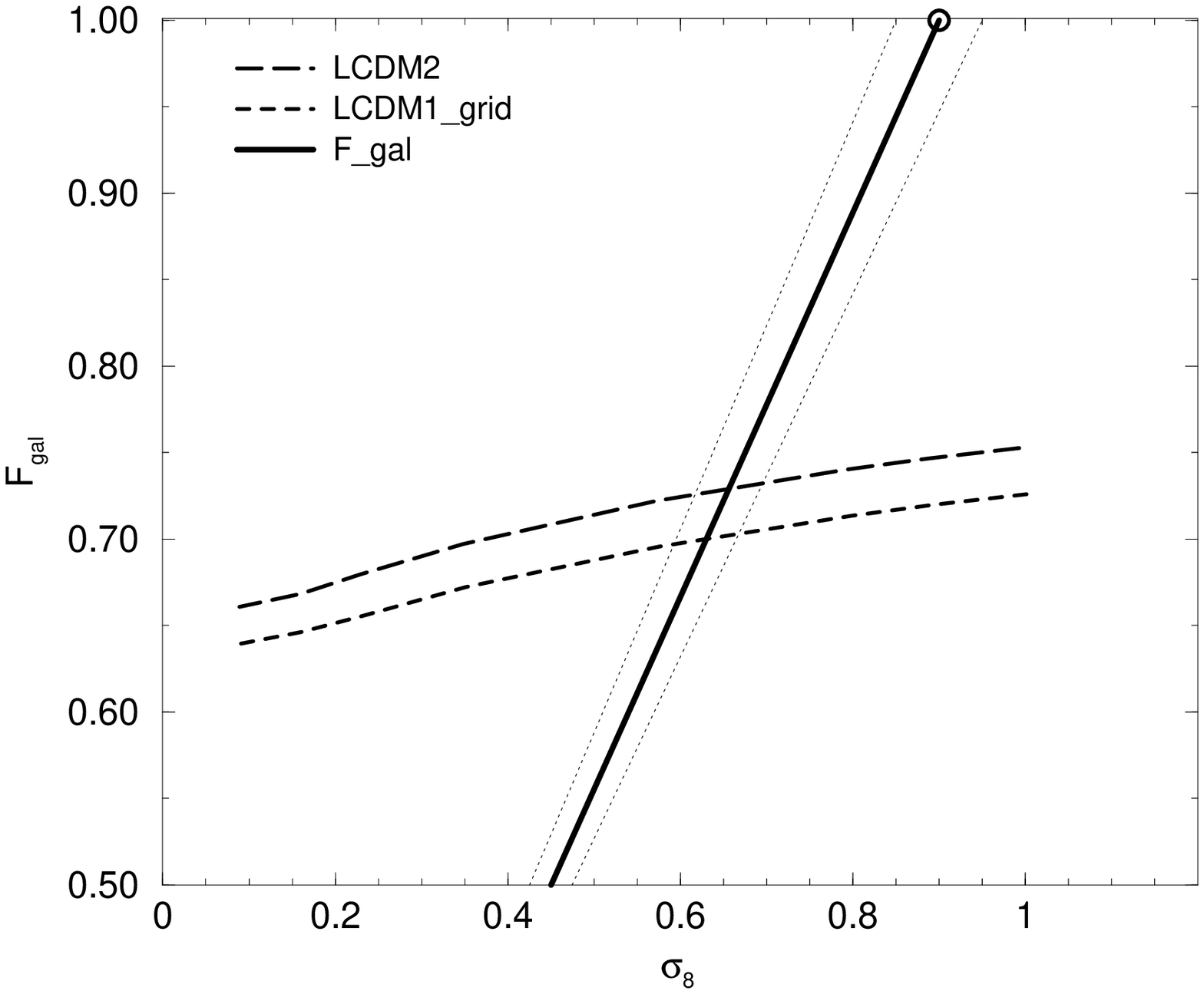} 
\includegraphics{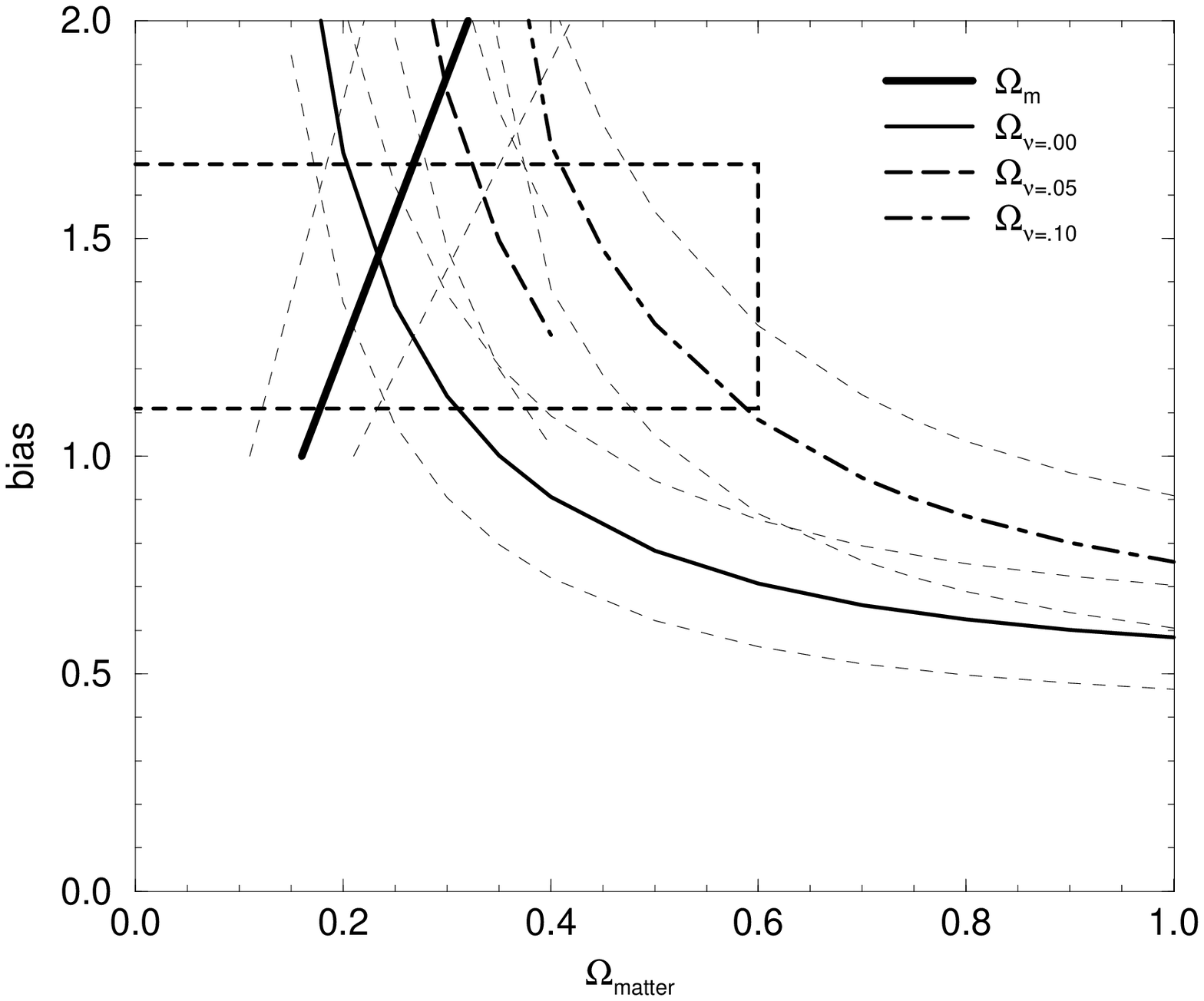}
\includegraphics{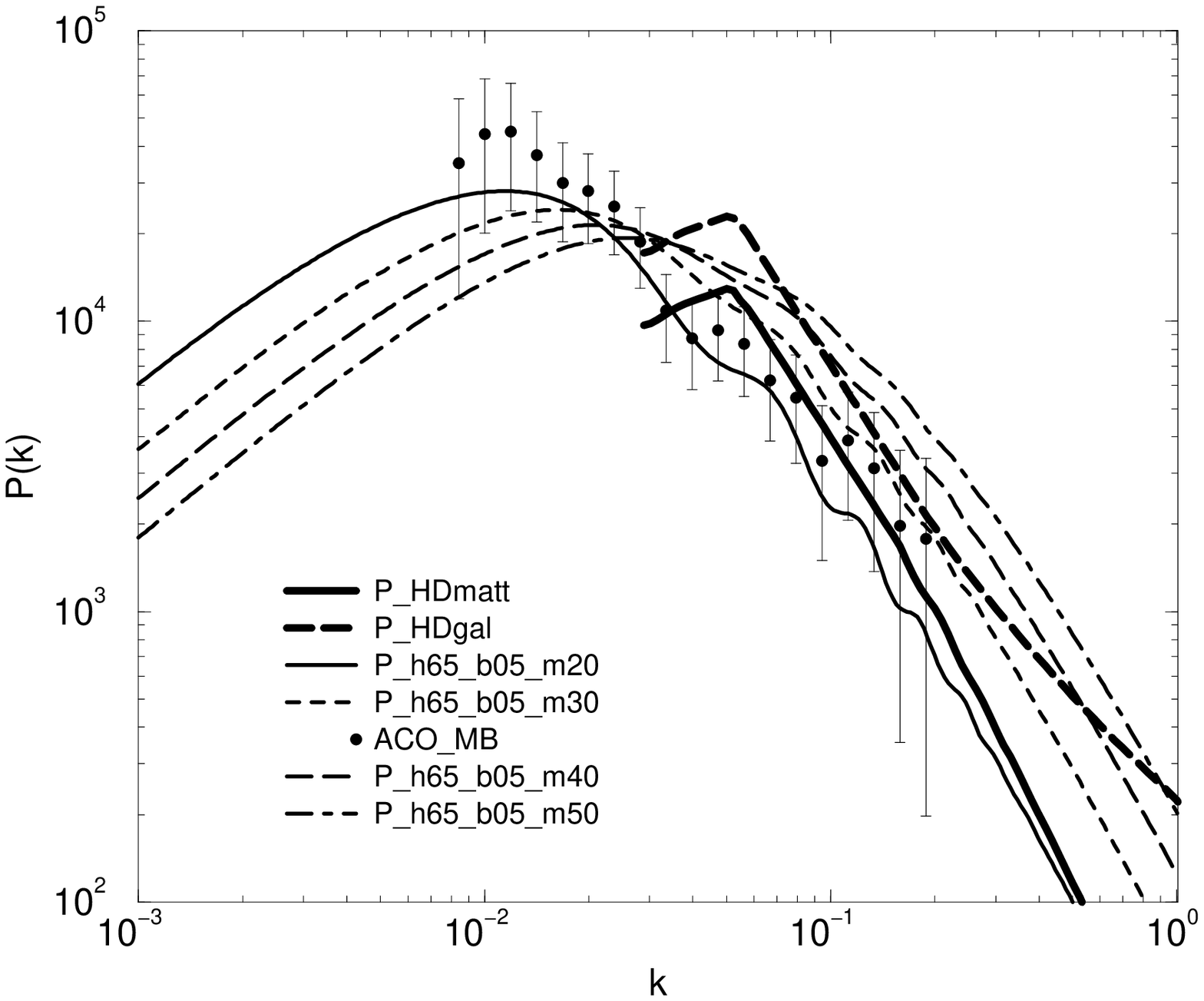} 
\includegraphics{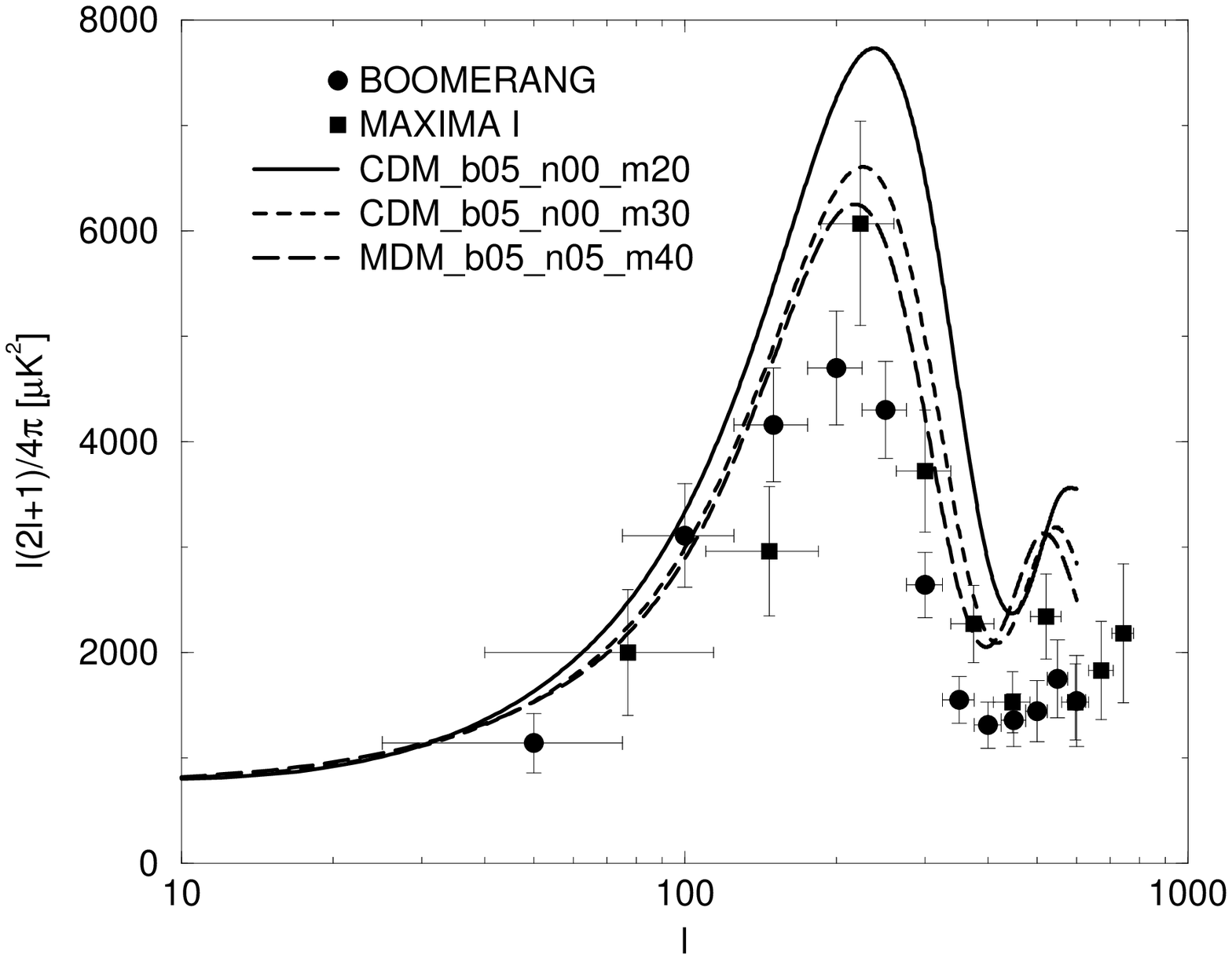}
\label{fig:model1}
\end{figure}

We express the epoch of simulations through the $\sigma_8$ parameter,
which was calculated by integrating the power spectrum of matter.  It
is related to the observed value of $(\sigma_8)_{gal}$ by the equation
(compare with eq.~(\ref{eq:sp2}, \ref{eq:bias}))
\begin{equation}
(\sigma_8)_{gal} = b_{gal} (\sigma_8)_m;
\label{eq:sigma}
\end{equation}
here we assume that $b_{gal} = b_c$.  This equation, and the observed
value of $(\sigma_8)_{gal}$, yields one equation between
$(\sigma_8)_m$ and $b_c$ (or $F_{gal}$); it is shown in the upper left
panel of Figure~\ref{fig:model1} by a bold line with error corridor.
The other equation is given by the growth of $F_{gal}$ with epoch. For
two LCDM models with density parameter $\Omega_m \approx 0.4$ the
growth of $F_{gal}$ is shown by dashed curves in the upper left panel
of Figure~\ref{fig:model1} \cite{e99b}.  By simultaneous solution of
both equations we found all three quantities of interest for the
present epoch: rms density fluctuations of matter $(\sigma_8)_m = 0.64
\pm 0.06$, the fraction of matter in the clustered population,
$F_{gal} = 0.70 \pm 0.09$, and the biasing parameter $b_{gal} = 1.4
\pm 0.1$.

The CMBFAST algorithm yields for every set of cosmological parameters
the $\sigma_8$ value for matter. It is calculated using the linear growth
model of density perturbations.  From observations we know this
parameter for galaxies, $(\sigma_8)_{gal}$.  Using eq.~(\ref{eq:sigma})
we can calculate the biasing parameter $b_{gal}$, needed to bring the
theoretical power spectrum of matter into agreement with the observed
power spectrum of galaxies.  This parameter must lie in the range
allowed by numerical simulations of the evolution of structure.
Results of calculations for a range of $\Omega_m$ are shown in the upper
right panel of Figure~\ref{fig:model1}, using the Hubble constant
$h=0.65$, baryon density $\Omega_b=0.05$, and HDM densities $\Omega_n =
0.00,~~0.05,~~0.10$.  The biasing parameter range shown in the Figure
is larger than expected from calculations described above; this range
corresponds to the maximum allowed range of the fraction of matter in
the clustered population expected from analytic estimates of the speed
of void evacuation.

Power spectra for LCDM models ($\Omega_n = 0$; $0.2 \leq \Omega_m \leq
0.5$) are shown in the lower left panel of Figure~\ref{fig:model1}.
We see that with increasing $\Omega_m$ the amplitude of the power
spectrum on small scales (and respective $\sigma_8$ values) increases,
so that for high $\Omega_m$ the amplitude of the matter power spectrum
exceeds the amplitude of the galaxy power spectrum. This leads to bias
parameter values $b \leq 1$.  Such values are unlikely since the
presence of matter in voids always increases the amplitude of the
galaxy power spectrum relative to the matter spectrum.  If other
constraints demand a higher matter density value, then the amplitude
of the matter power spectrum can be lowered by adding some amount of
HDM. However, supernova and cluster X-ray data exclude density values
higher than $\Omega_m \approx 0.4$; thus the possible amount of HDM is
limited.  The lower right panel of the Figure~\ref{fig:model1} shows
the angular spectrum of temperature anisotropies of CMB for different
values of the density parameter $\Omega_m$.  We see that a low
amplitude of the first Doppler peak of the CMB spectrum prefers a
higher $\Omega_m$ value: for small density values the amplitude is too
high. So a certain compromise is needed to satisfy all data.

\begin{figure}[ht]
\vspace*{6.0cm}
\caption{ Left: cluster mass distribution for LCDM models of various
density $\Omega_m$, with and without a Chung bump of amplitude
$a=0.5$. Right: cluster abundance of LCDM and MDM models of various density of
matter $\Omega_m$ and hot dark matter $\Omega_n$.  }
\includegraphics{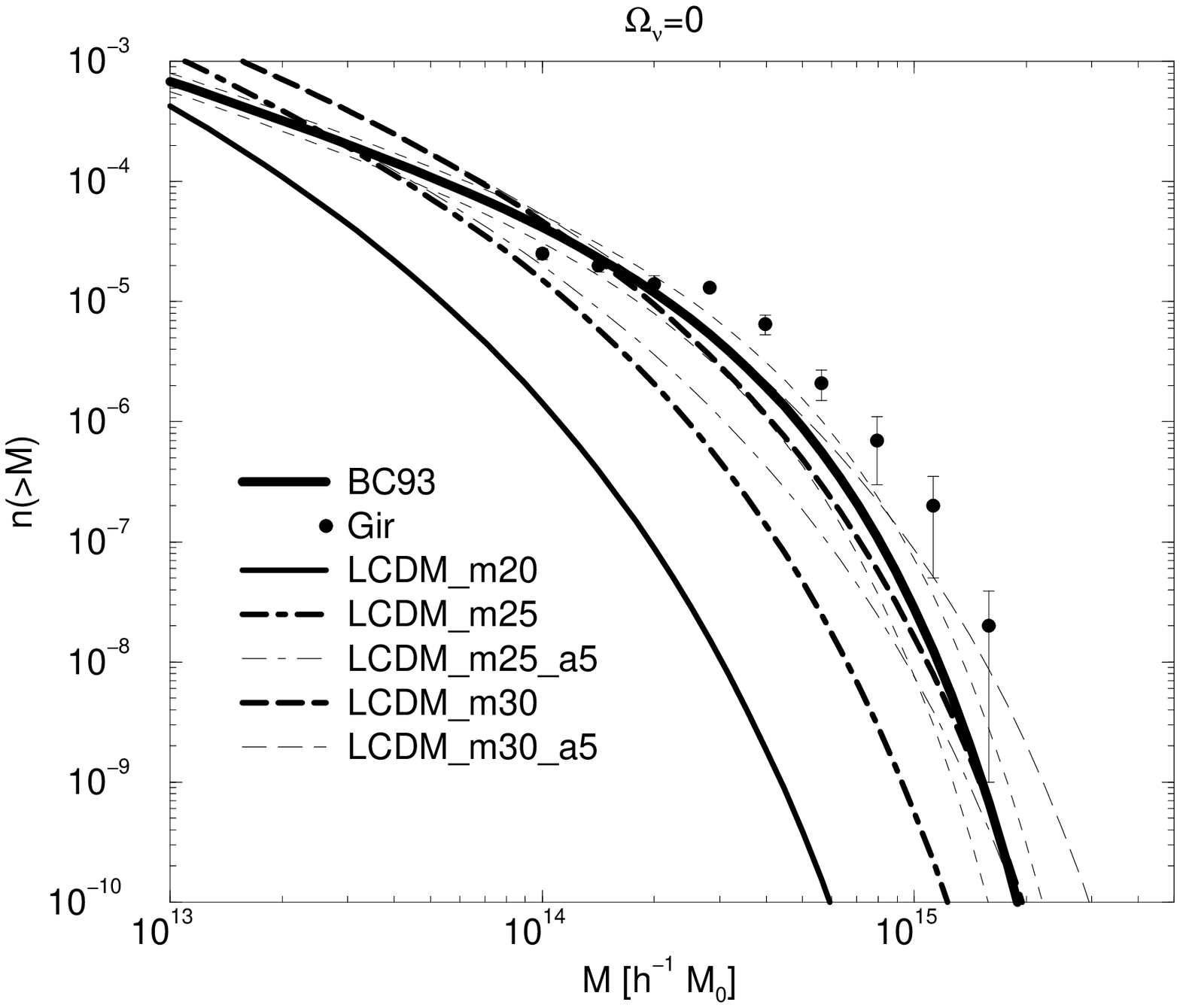} 
\includegraphics{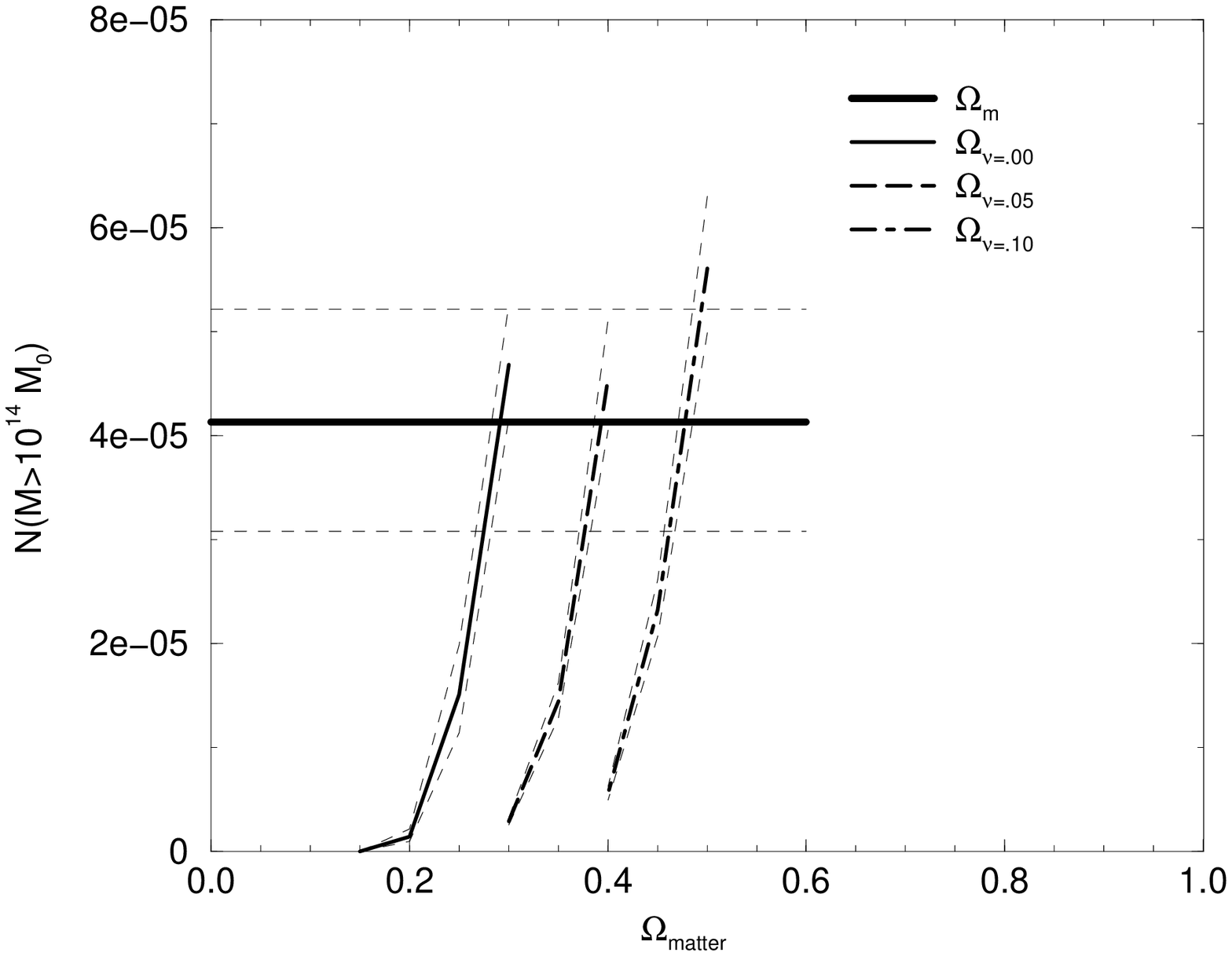}
\label{fig:mass}
\end{figure} 

The cluster mass distribution for LCDM models $0.2 \leq \Omega_m \leq
0.3$ is shown in the left panel of Figure~\ref{fig:mass}.  We see that
low-density models have a too low abundance of clusters over the whole
range of cluster masses. The best agreement with the observed cluster
abundance is obtained for a LCDM model with $\Omega_m = 0.3$, in good
agreement with direct data on matter density.  In this Figure we show
also the effect of a bump in the power spectrum, which is seen in the
observed power spectrum of galaxies and clusters \cite{e99a}.
Several modifications of the inflation scenario predict the formation
of a break or bump in the power spectrum. The influence of the break
suggested by Lesgourgues, Polarski and Starobinsky \cite{lps98} was
studied by Gramann and H\"utsi \cite{gh00}.  Another mechanism was
suggested by Chung \etal\  \cite{ckrt}. To investigate the latter case
we have used a value of $k_0 = 0.04$~\hmpc\ for the long wavenumber
end of the bump, and $a = 0.3 - 0.8$ for the amplitude parameter. Our
results show that such a bump only increases the abundance of very
massive clusters.  In the right panel of Figure~\ref{fig:mass} we show
the cluster abundance constraint for clusters of masses exceeding
$10^{14}$ solar masses; the curves are calculated for LCDM and MDM
models with $\Omega_n = 0.00,~~0.05,~~0.10$.  We see that the cluster
abundance criterion constrains the matter and HDM densities in a
rather narrow range.

\begin{figure}[ht]
\vspace*{11.5cm}
\caption{ Upper left: power spectra of a LCDM model with and without
Starobinsky modification.  Upper right: power spectra of MDM models
with and without Chung modification.  Lower left: cluster mass
distributions for MDM models with and without Chung modification.  Lower
right: angular power spectra of tilted MDM models with and without
Chung modification (amplitude parameter $a=0.3$).  
}
\includegraphics{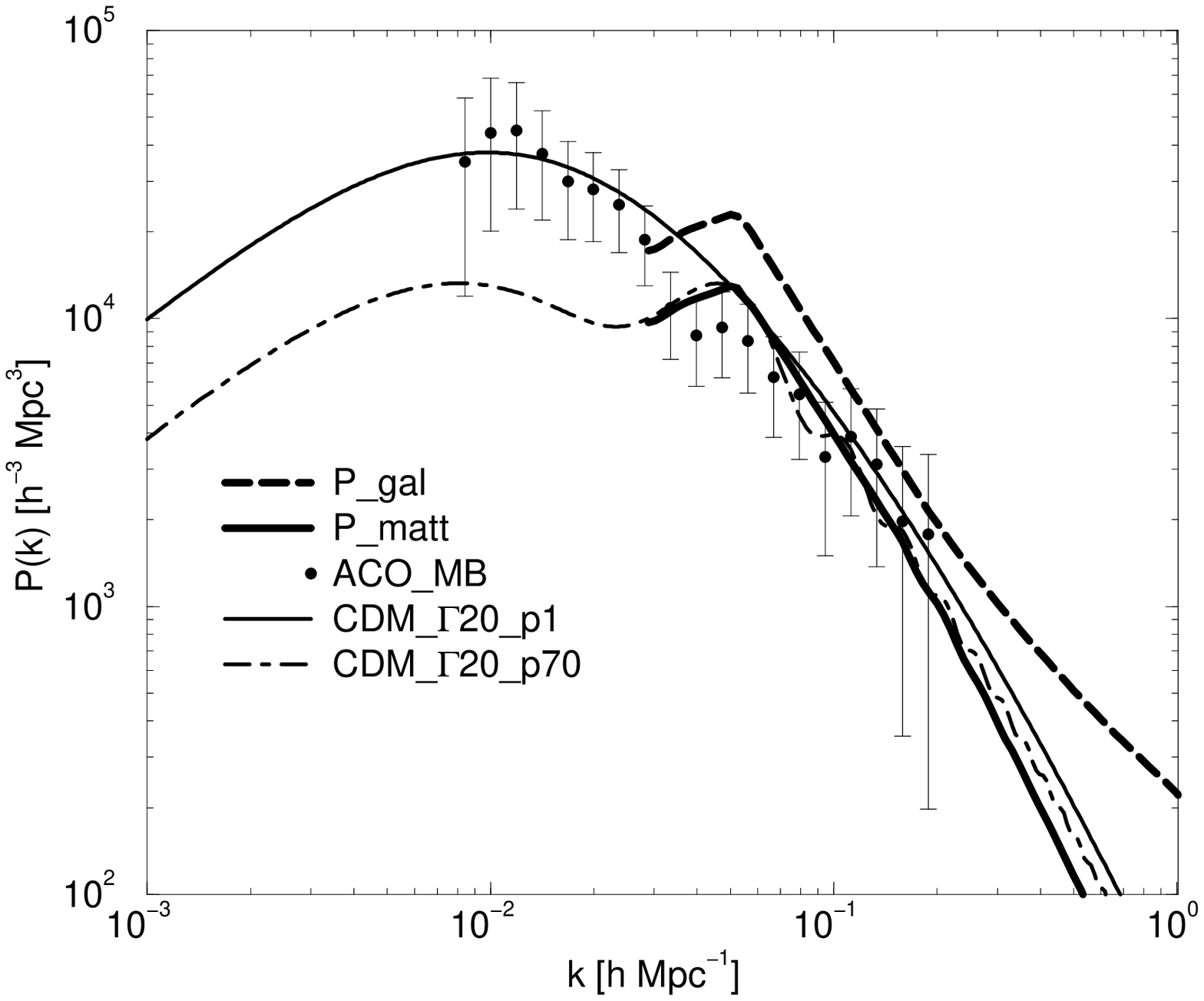} 
\includegraphics{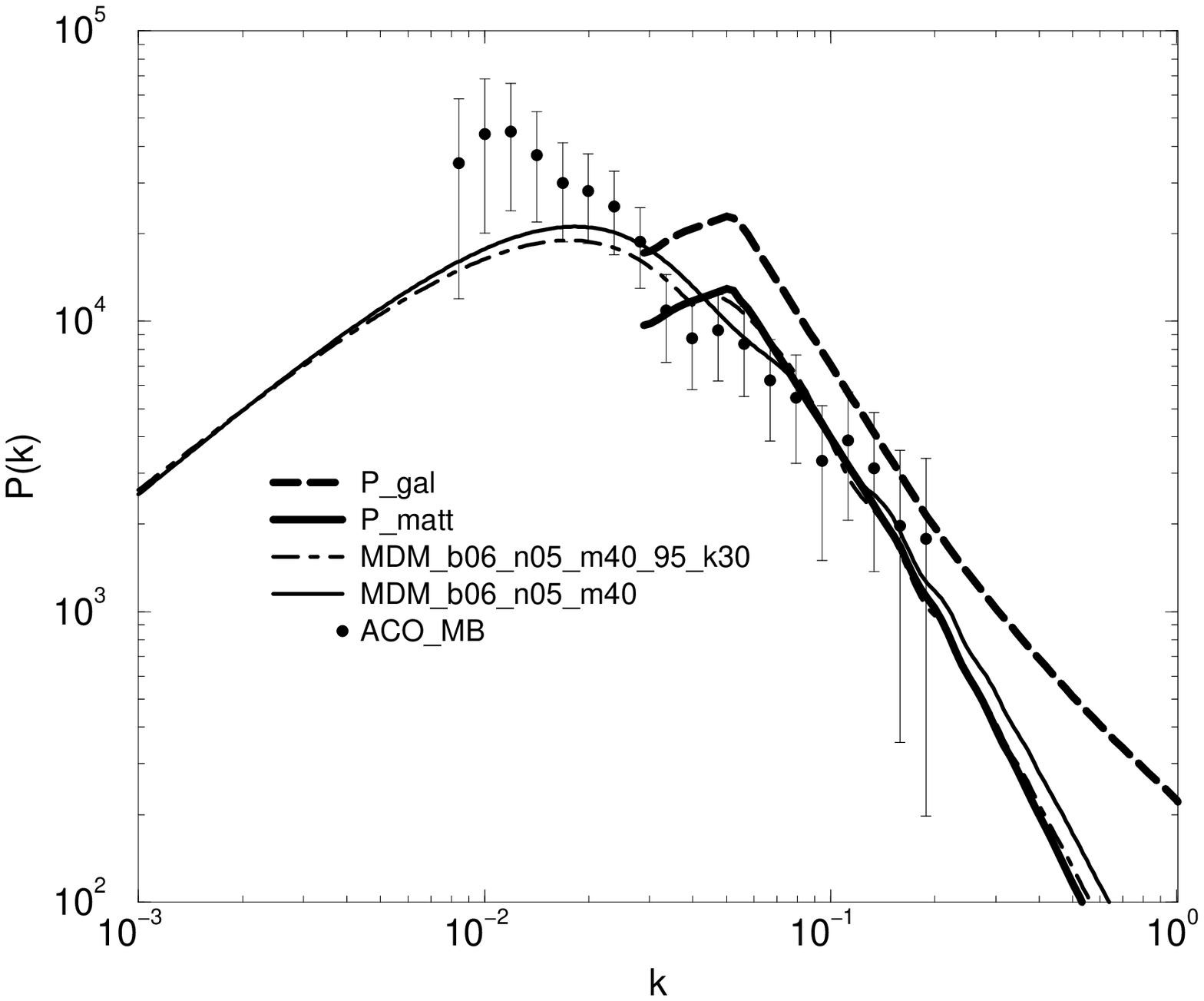}
\includegraphics{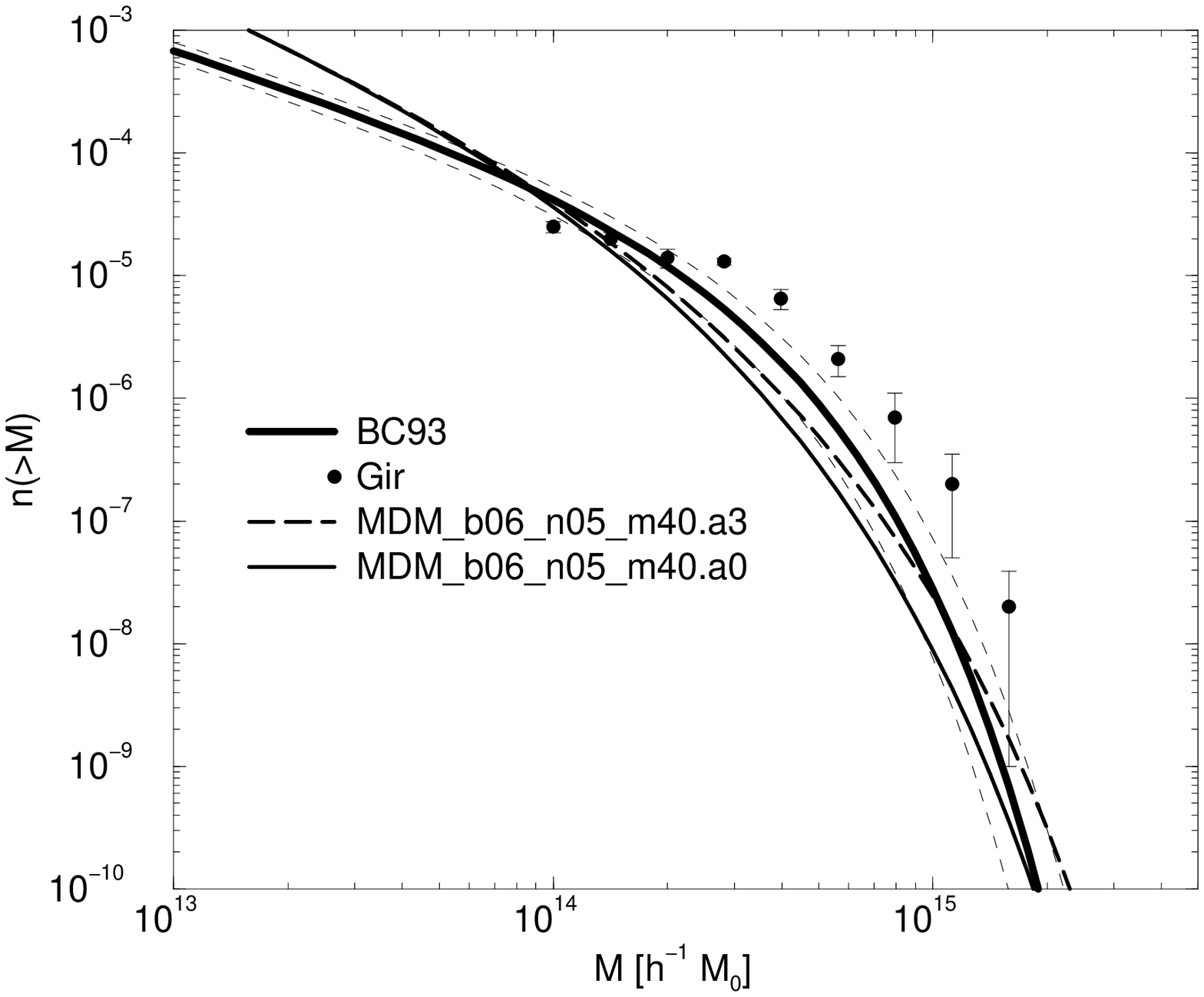} 
\includegraphics{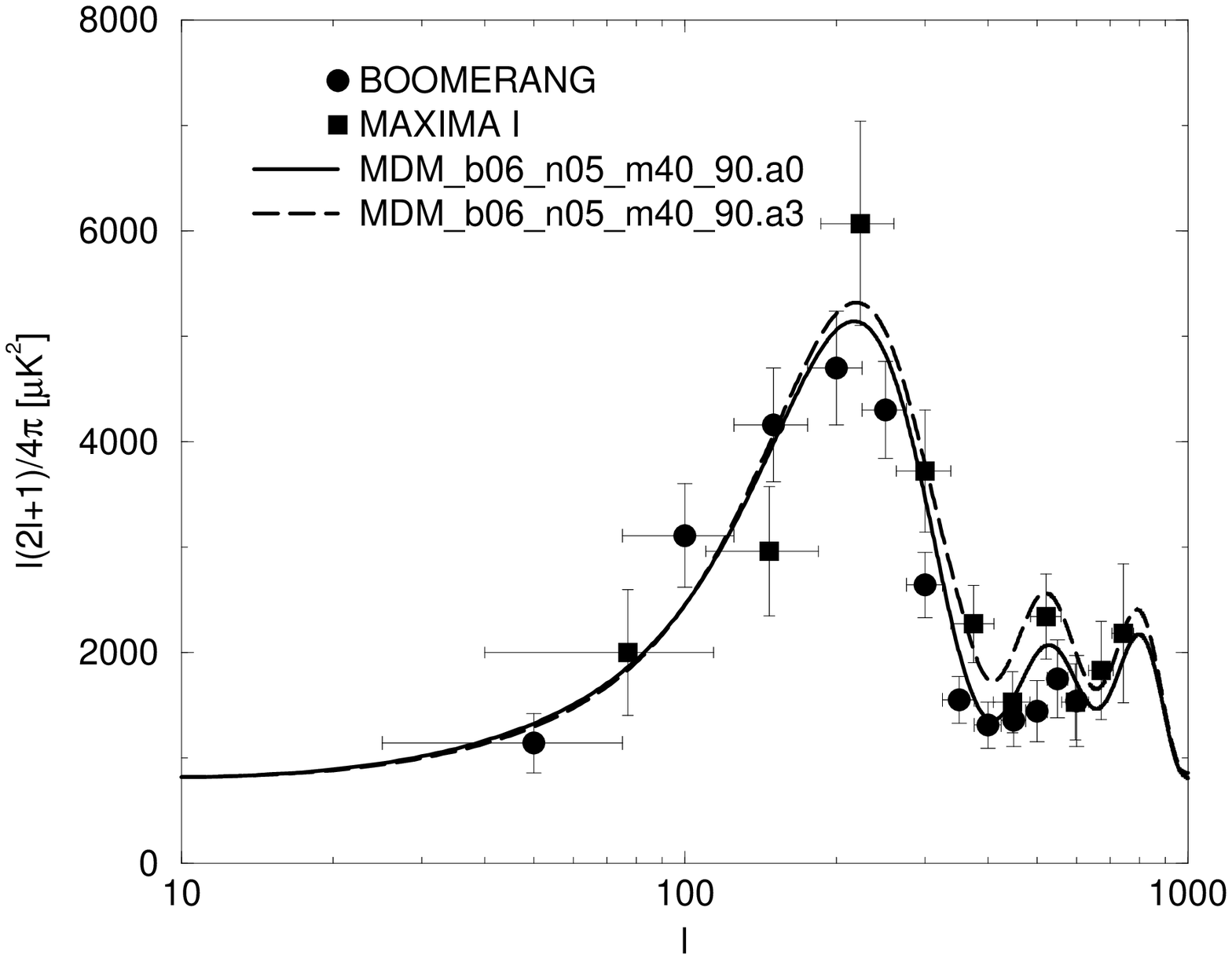}
\label{fig:model2}
\end{figure} 

The power spectra of LCDM models with and without the Starobinsky
break are shown in the upper left panel of Figure~\ref{fig:model2};
these models were calculated for the parameter $\Gamma = \Omega_m h =
0.20$.  In the case of the spectrum with a bump we have used MDM
models as a reference due to the need to decrease the amplitude of the
spectrum on small scales; these spectra are shown in the upper right
panel of Figure~\ref{fig:model2}.  Power spectra are compared with the
observed galaxy power spectrum \cite{e99a} and with the new cluster
power spectrum \cite{mb00}, reduced to the amplitude of the galaxy
power spectrum.  Also the matter power spectrum is shown, for which we
have used a biasing factor $b_c = 1.3$ \cite{e99b}.  We see that the
Starobinsky model reproduces well the matter power spectrum on small
and intermediate scales, but not the new data by Miller \& Batuski.
The modification by Chung \etal\ \cite{ckrt} with amplitude parameter
$a=0.3$ fits well all observational data.  The cluster mass
distribution for the Chung model is shown in the lower left panel of
Figure~\ref{fig:model2}, and the angular spectrum of CMB temperature
fluctuations in the lower right panel of Figure~\ref{fig:model2}.  In
order to fit simultaneously the galaxy power spectrum and the CMB
angular spectrum we have used a tilted MDM model with parameters
$n=0.90$, $\Omega_b = 0.06$, $\Omega_n = 0.05$, and $\Omega_m =0.4$.

BOOMERANG and MAXIMA I data have been used in a number of studies to
determine cosmological parameters \cite{b00,deb00,h00,tz00,wsp00}.  In
general, the agreement between various determinations is good;
however, some parameters differ.  There is a general trend to
interpret new CMB data in terms of a baryon fraction higher than
expected from the nucleosynthesis constrain; $h^2 \Omega_b = 0.03$.
Tegmark \& Zaldarriaga \cite{tz00} suggested a relatively high matter
density, $h^2 \Omega_m = 0.33$.  On the other hand, velocity data
suggest a relatively high amplitude of the power spectrum, $\sigma_8
\Omega_m^{0.6} = 0.54$, which in combination with distant supernova
data yields $\Omega_m = 0.28 \pm 0.10$, and $\sigma_8 = 1.17 \pm 0.2$
\cite{b00}.

Our analysis has shown that a high value of the density of matter,
$\Omega_m >0.4$, is difficult to reconcile with current data on
supernova and cluster abundances. Similarly, a high amplitude of the
matter power spectrum, $\sigma_8 >1$, seems fairly incompatible with
the observed amplitude of the galaxy power spectrum and reasonable
bias limits.  This conflict can be avoided using a tilted initial
power spectrum, and a MDM model with a moderate fraction of HDM, as
discussed above. The best models suggested so far have $0.3 \leq
\Omega_m \leq 0.4$, $0.90 \leq n \leq 0.95$, $0.60 \leq h \leq 0.70$,
$\Omega_n \leq 0.05$.  Matter density values lower than 0.3 are
strongly disfavoured by the cluster abundance constraint, and values
higher than 0.4 by all existing matter density estimates. This upper
limit of the matter density, in combination with the cluster abundance
and the amplitude of the power spectrum, yields an upper limit to the
density of hot dark matter.  We can consider this range of
cosmological parameters as a set which fits well all constraints.
This set of cosmological parameters is surprisingly close to the set
suggested by Ostriker \& Steinhardt \cite{os95}.  Now it is supported
by much more accurate observational data.

\ack {I thank M. Einasto, M. Gramann, V. M\"uller, A. Starobinsky,
E. Saar and E. Tago for fruitful collaboration and permission to use
our joint results in this review, and H. Andernach for suggestions to
improve the text.  This study was supported by the Estonian Science
Foundation grant 2625.  }

\end{document}